\documentclass[letterpaper,twocolumn,10pt]{article}
\usepackage{file}

\usepackage{tikz}
\usepackage{amsmath}
\usepackage{amsfonts}
\usepackage{filecontents}
\usepackage{algorithm}
\usepackage{algorithmic}
\usepackage{multicol}
\usepackage{multirow}
\usepackage{lipsum}
\usepackage{newfloat}
\usepackage{listings}
\usepackage{tabularx}
\usepackage{tcolorbox}
\usepackage{makecell}
\usepackage{enumitem}
\usepackage{amsmath}
\usepackage{subcaption}
\usepackage{booktabs}
\usepackage{array}
\usepackage{caption}
\usepackage{float}
\usepackage{textcomp}
\usepackage{newfloat}
\usepackage{listings}
\usepackage{subcaption} 
\usepackage{bm} 

\newtcolorbox{promptbox}{
  colback=gray!5!white, 
  colframe=gray!75!black, 
  rounded corners, 
  boxrule=0.3pt, 
  coltitle=black,
}

\begin{document}

\date{}

\title{\Large \bf LLM Embedding-based Attribution (LEA): Quantifying Source Contributions to Generative Model's Response for Vulnerability Analysis}

\author{
{\rm Reza Fayyazi}\\
Rochester Institute of Technology
\and
{\rm Michael Zuzak}\\
Rochester Institute of Technology
 \and
 {\rm Shanchieh Jay Yang}\\
Gonzaga University
} 

\maketitle

\begin{abstract}
Large Language Models (LLMs) are increasingly used for cybersecurity threat analysis, but their deployment in security-sensitive environments raises trust and safety concerns. With over 21,000 vulnerabilities disclosed in 2025, manual analysis is infeasible, making scalable and verifiable AI support critical. When querying LLMs, dealing with emerging vulnerabilities is challenging as they have a training cut-off date. While Retrieval-Augmented Generation (RAG) can inject up-to-date context to alleviate the cut-off date limitation, it remains unclear how much LLMs rely on retrieved evidence versus the model's internal knowledge, and whether the retrieved information is meaningful or even correct. This uncertainty could mislead security analysts, mis-prioritize patches, and increase security risks. Therefore, this work proposes \textit{LLM Embedding-based Attribution (LEA)} to analyze the generated responses for vulnerability exploitation analysis. More specifically, LEA quantifies the relative contribution of internal knowledge vs. retrieved content in  the generated responses. We evaluate LEA on 500 critical vulnerabilities disclosed between 2016 and 2025, across three RAG settings—valid, generic, and incorrect—using three state-of-the-art LLMs. Our results demonstrate LEA’s ability to detect clear distinctions between non-retrieval, generic-retrieval, and valid-retrieval scenarios with over 95\% accuracy on larger models. Finally, we demonstrate the limitations posed by incorrect retrieval of vulnerability information and raise a cautionary note to the cybersecurity community regarding the blind reliance on LLMs and RAG for vulnerability analysis. LEA offers security analysts with a metric to audit RAG-enhanced workflows, improving the transparent and trustworthy deployment of AI in cybersecurity threat analysis. \footnote{Code \& Data available at: \url{https://github.com/RezzFayyazi/LEA}}
\end{abstract}

\section{Introduction}
\label{sec/intro} 

Security professionals and vendors rely on threat analysis to defend against emerging attacks. However, the threat landscape is constantly evolving, with a sharp rise in both the volume and complexity of vulnerabilities \cite{okutan2023empirical}. Since 1999, the Common Vulnerabilities and Exposures (CVE) framework \cite{CVE} has cataloged more than 300,000 vulnerabilities, making it difficult to identify patterns and implement mitigations. Public databases like NIST NVD \cite{nvd2024} are essential for sharing this information, but rely heavily on manual analysis, which is slow and introduces delays in responding to emerging threats \cite{okutan2023empirical}. In this work, we consider a scenario in which a security analyst is reviewing their enterprise system for the presence of vulnerabilities to propose mitigation strategies in a timely manner.

Large Language Models (LLMs) have been widely adopted for cybersecurity threat analysis in enterprise systems \cite{khare2023understanding, deng2024pentestgpt, mitra2024localintel, perrina2023agir, chopra2024chatnvd}. They have been particularly effective in cyber threat analysis \cite{mitra2024localintel, perrina2023agir}, LLM-assisted attacks \cite{gupta2023chatgpt, deng2024pentestgpt}, and vulnerability detection \cite{cheshkov2023evaluation, chopra2024chatnvd}. LLMs are inherently constrained by their training data, which is time-gated because their knowledge reflects only what was available up to a fixed cutoff date. Meanwhile, thousands of new CVEs are disclosed each year, with over 21,000 in 2025 alone. This creates a persistent out-of-distribution problem: LLMs lack awareness of emerging vulnerabilities that appear after their training cutoff.

Therefore, as LLMs lack inherent awareness of post-training threats, modern deployments leverage Retrieval-Augmented Generation (RAG) \cite{borgeaud2022improving} to fetch up-to-date vulnerability data at inference time. 
However, as there is no foundation model that contains explicit knowledge of all CVEs, a RAG pipeline retrieves vulnerability descriptions and metadata to ground the model response. This creates two possibilities: (1) the retrieved content is relevant and the model provides a grounded, helpful answer; or (2) the retrieval is irrelevant, incomplete, or incorrect, which leads the model to hallucinate or fabricate mitigation steps. Therefore, as LLMs are adopted into threat analysis pipelines, security analysts must address the risk of flawed LLM generation, such as inaccurate vulnerability descriptions or improper mitigation procedures, to avoid further compromise of system integrity. 

This raises a critical challenge for security analysts: when an LLM generates a response to a vulnerability, to what extent is that answer grounded in retrieved evidence versus inferred from the model’s pre-trained knowledge? In high-stakes environments where inaccurate mitigation advice can jeopardize system integrity, understanding this dependency is vital. Specifically, we seek to quantify the `dependence distribution' of an LLM’s output—how much of it is shaped by the user query, the retrieved context, or the model’s internal parameters. As LLMs become embedded in security workflows, developing methods to trace, interpret, and audit this interplay is essential for ensuring trustworthy and accountable decision-making.

One way to quantify this can be with token-level probability differences. However, as vulnerability jargon often has multiple forms for the same concept, e.g., `Cross-Site Scripting' vs. its abbreviation `XSS' or `SQL Injection' vs. `SQLi'. An LLM can assign different token-level probabilities even though these pairs convey identical meaning. For this reason, the vulnerability analysis problem is unique from broader retrieval tasks. To address this problem, we rely on the observation that LLMs exhibit substantial linear dependency, drawn from prior work on parameter efficient fine-tuning (PEFT) methods, particularly LoRA \cite{hu2021lora}.
From a theoretical perspective, the embedding vectors inhabit nearly the same subspace, so one can be expressed as a linear combination of the other. The \textit{rank} of a matrix reflects the number of linearly independent vectors it contains, offering a measure of the \textit{unique} or \textit{non-redundant} information present. Therefore, if a vector can be expressed as a linear combination of others, it does not contribute new information, indicating redundancy. In contrast, a set of linearly independent vectors implies that each vector adds as a distinct value to the representation space. 

Built upon the above insights, we introduce a novel, explainable metric called \textbf{LLM Embedding-based Attribution (LEA)}, based on the linear dependence of hidden state representations within LLMs. By measuring the rank of a matrix constructed from both retrieved-augmented representations and the latent hidden states of the LLM, we can quantitatively assess the relative contribution of each source to the final output, by showcasing the dependence distribution of the generated response on the input query, retrieved context, and internal knowledge. 
We curated 500 critical and high severity CVEs from 2016 to 2025, and conducted an analysis of LEA across the layers of three instruction-tuned LLMs of varying sizes: \textit{Gemma-3-27B-IT} \cite{gemma3}, \textit{Mistral-Small-24B-Instruct-2501} \cite{mistral}, \cite{deepseek8b}, and \textit{LLaMA-3.2-3B-Instruct} \cite{llama32} to show generalizability. 

We test LEA with the baseline of non-retrieval generations against three different RAG scenarios, namely valid, generic, and incorrect retrieval.  In the valid retrieval case, we assume that the LLM retrieved only the most relevant and verified information from the NVD website \cite{nvd2024}. This is intended to further compare the LEA distribution with cases where the retrieval contains generic/incorrect content. For the generic retrieval case, we assume that the LLM does not know about a CVE and tries to retrieve generic information for what a CVE is (e.g., NVD homepage). Finally, in the incorrect retrieval scenario, we assume that the LLM retrieved the wrong information regarding a vulnerability (i.e., another CVE's description).

Our findings indicate that LLMs exhibit strong context independency as progresses through the transformer layers, which enables the construction of LEA to trace back to early layers to examine the linear dependence of the generated tokens against the retrieved context vs. LLM's internal knowledge. Our results reveal a common trend across all evaluated models: as long as the generated response captures the factual, verified content from RAG, LEA shows a relatively balanced distribution between the retrieved context and the LLM's internal knowledge.  
Conversely, when the model disregards verified RAG embeddings and generates a response with little to no alignment with the retrieved context, LEA exhibits an abnormal distribution in which few output tokens in the generated response can be attributed with the retrieved context. We verify LEA’s effectiveness in capturing and modeling the dependence distribution of generated tokens for vulnerability analysis. Our empirical results further demonstrate LEA’s ability to clearly differentiate between non-retrieval, generic-retrieval, and valid-retrieval scenarios with over 95\% accuracy on larger models. This establishes LEA's utility in evaluating responses that deviate from trusted intelligence sources.

Furthermore, through an analysis of LEA distributions on CVE data spanning 2016 to 2025, we expose the limitations of relying solely on LLM memorization, as there is a pool of over 300,000 disclosed CVEs. This indicates the inherent constraints of relying solely on foundational (internal) knowledge for comprehensive and up-to-date vulnerability analysis. Finally, we demonstrate the limitations posed by incorrect retrieval of vulnerability information and raise a cautionary note to the cybersecurity community regarding the blind reliance on LLMs for vulnerability analysis. By providing an auditable, layer-aware attribution metric, LEA helps Security Operations Centers (SOC) and threat-intel teams assess when RAG outputs can be trusted, when human verification is required, and where retrieval or provenance controls must be tightened. We summarize our contributions as follows:

\begin{itemize}

    \item {\bf Explainable Metric:} A novel metric, \textbf{LEA}, is developed to reveal how LLM generated responses are attributed across the input query, retrieved context, and foundational knowledge.
    \item {\bf Benchmarking LEA for Verifiable Vulnerability Insights:} 
    We evaluate LEA’s effectiveness in verifying LLM-generated insights for vulnerability analysis by establishing the expected LEA distribution when aligned with trusted ground-truth sources, using 500 CVEs spanning a decade.
    \item {\bf Discriminating Informative vs. Generic Vulnerability Explanations:} 
    We demonstrate that LEA reliably distinguishes between informative and non-informative responses across non-retrieval, generic-retrieval, and valid-retrieval settings with over 95\% detection accuracy for larger models.
    \item {\bf Pattern Learning over Memorization of CVEs:}
    LEA reveals that LLM responses on structured entities such as CVE-IDs emerge from generalization patterns rather than direct memorization, raising critical concerns about reliability and the risks of blindly deploying LLMs for vulnerability analysis.

\end{itemize}

\begin{figure*}[t]
\centering
\includegraphics[scale=0.22]{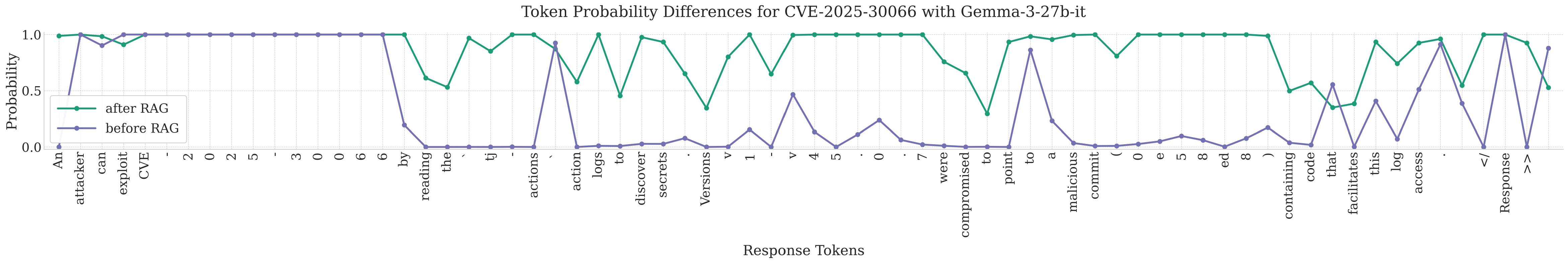}
\caption{Probability shifts in the RAG-generated response of a CVE before and after including the retrieved content.}
\label{fig:explain_tokens}
\end{figure*}

\begin{figure}[t]
\centering
\includegraphics[scale=0.28]{./figures/lea_example_2.pdf}
\caption{LEA's distribution for a vulnerability.}
\label{fig:example}
\end{figure}

\section{Preliminaries}
\label{sec/related_works}

\subsection{Challenges of Content Retrieval for LLMs}

The knowledge cutoff of LLMs makes them prone to hallucinations when encountering new or evolving information (e.g., vulnerabilities) \cite{huang2023survey, rawte2023troubling}. 
Retrieval-augmented generation (RAG) \cite{borgeaud2022improving} mitigates this by injecting up-to-date external context at inference, and recent variants \cite{jeong2024adaptive, selfrag, kim2024sure} further enhance precision and robustness.
However, hallucinations may still arise, particularly when queries are ambiguous, retrieved documents are noisy or excessive, or the underlying corpus is of low quality \cite{zhang2025hallucination}. These limitations emphasize the need to quantify the dependence of a model on the retrieved context versus its internal knowledge, allowing better control, interpretability, and trust in generated outputs.

\subsection{LLMs in Vulnerability Analysis}

Prior work has explored the integration of LLMs into vulnerability analysis workflows \cite{khare2023understanding, cheshkov2023evaluation, du2024vul, chopra2024chatnvd}. Cheshkov et al. \cite{cheshkov2023evaluation} applied GPT-based models to detect vulnerabilities in Java code, showing that LLMs often struggle with identifying vulnerable patterns. In contrast, Khare et al. \cite{khare2023understanding} demonstrated that LLMs can outperform deep learning models in vulnerability detection when guided by well-crafted prompt strategies, which indicates the critical importance of prompt engineering in effectively leveraging LLMs for cybersecurity tasks.

Extending these efforts, Du et al. \cite{du2024vul} introduced Vul-RAG, a retrieval-augmented framework that constructs a multi-dimensional knowledge base from historical CVE reports and integrates it with LLMs to improve contextual understanding during vulnerability analysis. 
While Vul-RAG grounds responses in structured, real-world data, its dependence on manual curation and knowledge engineering limits its scalability and effectiveness in real-time or automated threat detection scenarios. 
ChatNVD \cite{chopra2024chatnvd} uses NVD data and a GPT-4 variant to deliver compact CVE summaries, helping analysts grasp each vulnerability’s nature, exploitability, and impact. 
Despite these promising advances, none of these works quantified how much LLM outputs depend on retrieved knowledge versus internal representations, an important gap in explainability and trust for cybersecurity decision-making.

\section{Linear Independence for Source Attribution}

To better understand the influence of RAG on the LLM generations, we analyze the difference in token-level probabilities before and after RAG. This analysis reveals how RAG affects the model’s confidence in generating specific tokens. To do this, we get the Softmax probabilities for the response tokens before and after including the retrieved content. Figure \ref{fig:explain_tokens} shows how token-level probabilities in the generated response shift when the retrieved content is provided. Notably, the probabilities of key tokens related to \textit{CVE-2025-30066} exhibit substantial changes, which indicates the influence of the retrieved content on the model’s output. 
While informative, token-level probability changes alone do not quantify how much the model relies on retrieved information, the underlying goal of this work. One limitation arises from the linguistic variability inherent in cybersecurity jargon. For instance, terms such as ``malicious commit'' and ``backdoor commit'' may be semantically equivalent, yet the model may assign distinct token probabilities to each. This semantic ambiguity complicates efforts to attribute generation influence based purely on token probability shifts.

\begin{figure*}[t]
\centering
\includegraphics[scale=0.2]{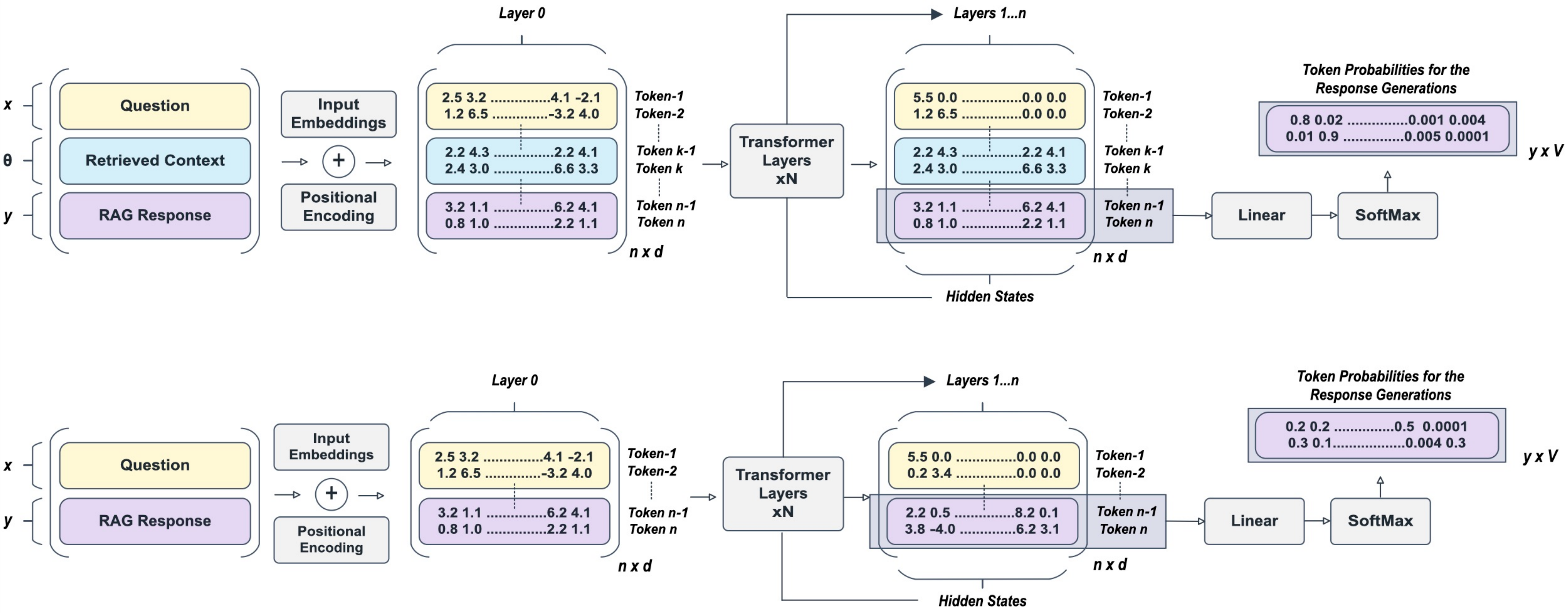}
\caption{The step-by-step of getting the hidden state progression and probability differences.}
\label{fig:probs_process}
\end{figure*}

To overcome this challenge, we introduce an interpretable approach grounded in the internal structure of the model's hidden state representations. Prior work has shown that transformer models often exhibit redundancy in their weight matrices and hidden states, with many directions in these high-dimensional spaces contributing little unique information \cite{hu2021lora, ma2023llm,ding2025sliding}. One notable method that capitalizes on this property is LoRA \cite{hu2021lora}. LoRA has gained widespread adoption for fine-tuning LLMs as it achieves performance comparable to full fine-tuning while requiring updates to only a few million parameters.
Recent research in attention head pruning has shown that a substantial number of attention heads can be removed with minimal impact on performance \cite{ma2023llm, ding2025sliding}. Pruning techniques aim to remove redundant parameters or structures, such as attention heads or neurons, based on the insight that many components in LLMs learn features that are linear combinations of others, and thus do not contribute uniquely to model output. 

This insight motivates the use of matrix ``rank'' as a metric to assess linear independence and information diversity across hidden-state matrices.
The rank of a hidden-state matrix quantifies the number of linearly independent vectors, effectively capturing the dimensionality of the informative (i.e., task-relevant) subspace in a given layer. A lower rank implies redundancy or strong dependence among token representations, while a higher rank suggests that the tokens are contributing distinct information to the representation space. By computing the rank at each transformer layer, we gain visibility into how token interactions change with depth, and how the presence of RAG-derived context modulates these interactions. 
This leads us to consider how the concept can be applied in the analysis of vulnerabilities to detect which token generations in the responses were from the retrieved content. In this setting, rank-based analysis can help determine whether the model's response about a CVE (e.g., its exploitability and impact) is primarily grounded in retrieved content or generated using internal knowledge.

Figure \ref{fig:example} shows the dependence distribution produced by our proposed metric (LEA) for \textit{CVE-2025-30066}. A full mathematical definition for LEA can be found in the next section. The resulting LEA distribution indicates that for this example, 26\% of the generated content is attributed to the model’s internal (or foundation) knowledge, while 13\% and 61\% are influenced by the input query and the retrieved documents, respectively. 
This dependence distribution is useful for identifying low-quality retrieval and hallucinations in downstream cybersecurity tasks. If the dependence distribution with retrieved tokens closely mirrors that seen in the absence of retrieval, it could indicate that the model possesses internal insights about the CVE and retains semantically relevant associations from its foundational knowledge. In contrast, when the model overlooks verified RAG embeddings and produces responses with minimal focus on the retrieved content, it could indicate hallucination or a generic generation (i.e., an uninformative response). In the context of vulnerability analysis, a poorly aligned response can result in misleading interpretations, causing analysts to mis-prioritize patches and elevate security risks. Therefore, detecting and filtering poorly grounded responses before dissemination is necessary to ensure the reliability of AI-assisted cybersecurity workflows.

\section{LEA: Mathematical Foundations}
\label{sec: math}
To demonstrate how the proposed LEA metric is derived, we first need to discuss how text progresses through a transformer architecture \cite{vaswani2017attention}. In a transformer, every layer ($\ell$) re-encodes the entire prompt, allowing each token (\textit{i})’s representation to depend on all other tokens. Consequently, the hidden states evolve from locally grounded embeddings ($\mathbf{e}_i$) to highly contextual vectors $\mathbf{x}_i^{(l)}$ (i.e., hidden states) that capture syntax, semantics, and long-range dependencies of the prompt. More specifically, a prompt is converted into an embedding matrix representing the tokens in the prompt, with the dimension of $\mathbb{R}^{L\times d}$, where \textit{L} is the number of tokens and \textit{d} is the dimension of the model. 
Let a prompt be tokenized in the ordered sequence $T=(t_1,t_2,\dots,t_L)$ of length~$L$.  
\begin{enumerate}[label=\arabic*)]
\item \emph{Token embedding.}  
      Each discrete token $t_i$ is mapped to a continuous vector via an embedding
      matrix $E\in\mathbb{R}^{|\mathcal{V}|\times d}$:
      \begin{equation}
          \mathbf{e}_i \;=\; E[t_i]\in\mathbb{R}^{d},
      \end{equation}

      where $d$ is the model (hidden) dimension and $\mathcal{V}$ is the vocabulary.

\item \emph{Positional encoding.}  
      Because the self-attention mechanism is permutation invariant, positional
      information must be injected.  
      For position~$i$ we add a vector $\mathbf{p}_i\in\mathbb{R}^{d}$ (either a fixed sinusoidal
      code or a learned embedding):

      \vspace*{-8pt}
      
    \begin{equation}
                \mathbf{x}_i^{(0)}
              \;=\;
              \mathbf{e}_i
              \;+\;
              \mathbf{p}_i,
              \qquad i=1,\dots,L.  
    \end{equation}

\vspace*{-2pt}

      The matrix $X^{(0)}=[\mathbf{x}_1^{(0)},\dots,\mathbf{x}_L^{(0)}]\in\mathbb{R}^{L\times d}$
      constitutes the \emph{layer-0 hidden state}.  It encodes \emph{both} lexical identity
      and position.

\item \emph{Stack of Transformer blocks.}  
      For $\ell=1,\dots,N$ (where $N$ is the number of layers) the hidden state is updated by a
      residual sublayer comprising multi-head self-attention (MHA), feed-forward network
      (FFN), and layer normalisation (LayerNorm):

      \vspace*{-12pt}
      
    \begin{equation}
    \tilde{X}^{(\ell)} = X^{(\ell-1)} + \mathrm{MHA}\!\Bigl(\mathrm{LayerNorm}(X^{(\ell-1)})\Bigr)
    \end{equation}

    \vspace*{-8pt}
    
    \begin{equation}
    X^{(\ell)} = \tilde{X}^{(\ell)} + \mathrm{FFN}\!\Bigl(\mathrm{LayerNorm}(\tilde{X}^{(\ell)})\Bigr)
    \end{equation}

      yielding $X^{(\ell)}\in\mathbb{R}^{L\times d}$, the \emph{layer-$\ell$ hidden state}.  
      Each row $\mathbf{x}_i^{(\ell)}$ is a context-dependent representation of the
      $i$-th token, refined through $\ell$ rounds of self-attention and nonlinear mixing.

\item \emph{Output (logits).}  
      After the final layer $N$, the hidden state $X^{(N)}$ is projected back to
      the vocabulary space through a linear map (often the transpose of $E$)
      followed by a soft-max:

      \vspace*{-12pt}
      
    \begin{equation}
    Z = X^{(N)} W_{\text{out}} + \mathbf{b}_{\text{out}} \;\in\; \mathbb{R}^{n \times |\mathcal{V}|}
    \end{equation}

    \vspace*{-12pt}
    
    \begin{equation}
    P = \operatorname{Softmax}(Z)
    \end{equation}

    yielding for every position $i$ a probability distribution
    $P_{i,:}$ over the next token.

\end{enumerate}

Overall, as embedding vectors propagate through the model’s transformer blocks, self-attention and feed-forward layers iteratively refine these vectors, producing a sequence of hidden states, one per token, that captures progressively richer semantic and syntactic context. Since decoder-only LLMs are trained with a causal mask, the hidden state at position $t$ is conditioned exclusively on tokens $1, \dots, t-1$, ensuring the model’s predictions remain autoregressive. In Figure \ref{fig:probs_process}, we show the steps for getting hidden states and probability differences with or without using RAG.

Now, we discuss the linear independence of these hidden state vectors. A set of vectors is linearly independent if no vector in the set can be expressed as a linear combination of the other vectors in the set. 
Let $\{v_1,v_2,\dots,v_n\}$ be a set of vectors in a vector space $V$.  
The vector $v_i$ is said to be \emph{linearly independent of the remaining vectors} if no combinations of scalars $\{\alpha_j, \forall j\neq i\}$ can produce $v_i$ except the obvious case.

\vspace*{-4pt}

\begin{equation}
    \sum_{\substack{j=1\\ j\neq i}}^{n} \alpha_j\,v_j \;\neq\; v_i
    \;\;\textrm{except}\;\;
    \alpha_j = 0
    \quad\forall j\neq i.
\end{equation}

The \textit{rank} of a matrix is how many vectors are needed to describe the original matrix. For \(X \in \mathbb{F}^{m \times n}\), the \emph{rank} of \(X\), is the dimension of its row-space:

\vspace*{-8pt}

\begin{equation}
    \operatorname{row}(X) = \operatorname{span}\{X_{1,:}, \dots, X_{m,:}\} \subseteq \mathbb{F}^{m}
\end{equation}

 In other words, \(\operatorname{rank}(X)\) counts the number of \emph{fundamental directions} needed to reconstruct every row (or, equivalently, every column) of \(X\).
        Any additional vector lying in the span of those directions adds no new information.
        
\vspace{2pt}

\noindent \textbf{Duplicate information.}
                If \(\operatorname{rank}(X)=r<m\), then exactly \(m-r\) rows are redundant.  
                Each redundant row can be written as a linear combination of \(r\) independent rows:

\vspace*{-4pt}

\begin{equation}
    X_{j,:} \;=\; \sum_{i=1}^{r}\alpha_{i}\,X_{k_{i},:}
\end{equation}

\noindent where \(k_{1},\dots,k_{r}\) index linearly independent rows and \(\alpha_{1},\dots,\alpha_{r}\in\mathbb{F}\).

Therefore, this concept can be applied in the context of LLMs, as each token has a vector representation. This structural property opens a pathway for fine-grained attribution analysis: by examining how hidden states evolve in the presence of retrieved context regarding new vulnerabilities, we can quantify the extent to which specific token generations in the model’s output are influenced by the retrieved evidence. 
To do so, we first concatenate the original question ($x$) with the RAG-generated response ($y$) to form the combined sequence $xy$, and we compare with adding the retrieved context ($\theta$) to get the sequence $x\theta y$. We then evaluate the contextual dependency of each token within this sequence.  Specifically, we compare the token-level dependency before and after incorporating $\theta$ (i.e., the retrieved content). Table \ref{tab:cheatsheet} compiles all symbol and term definitions used throughout the paper.

\begin{table}[t]
  \centering
  \footnotesize
  \caption{Brief description of the used terms in the paper}
  \label{tab:cheatsheet}

  \begin{tabularx}{\columnwidth}{@{} >{}m{0.09\columnwidth}<{} X @{}}
    \toprule
    \textbf{Term} & \textbf{Description}\\
    \midrule
    $x$     & input query\\
    $\theta$         & retrieved context (RAG)\\
    $y$              & generated response\\
    $xy$             & concatenation of the question and the response\\
    $x\theta y$      & concatenation of the question, retrieved context, and the response\\
    \bottomrule
  \end{tabularx}
\end{table}

Now, to see the progression of token independence across layers in different LLMs, we compute the rank of the hidden state matrices for both $xy$ and $x\theta y$ sequences. This rank quantifies the dimension of the span of token representations (i.e., how many are linearly independent). Based on our analysis of multiple CVEs, we observed that as the model progresses through the intermediate layers, a shift occurs: tokens are increasingly treated as independent variables, and the trend continues until the final layers. 

In Table \ref{table:rank_evolution_combined}, we demonstrate the change of \textit{rank} across the layers of the LLMs, with and without RAG, on two different CVEs. As can be seen, each model increasingly treats each token as an independent variable as depth grows, with a notable degree of independence already present in the earliest layers. 
This suggests that deeper layers re-encode representations in ways that abstract away initial inter-token dependencies. Our additional tests on various other CVEs further confirmed the consistency of this trend. We hypothesize that this behavior stems from the attention mechanism itself, which is designed to assign independent attention weights to each token. This increased independence could arise from the attention mechanism’s independent token weighting, which treats tokens in isolation and explains why LLMs must be so large to capture complex interdependence. 

Therefore, to capture dependency within tokens, we need to trace back to earlier layers. More specifically, we observe that layer-0, which includes both token embeddings and positional encodings, provides a signal of how dependent the model is on the input prompt. At this layer, the model's representations reflect meaningful contextual relationships between tokens, which indicates that layer-0 can be especially useful for discerning how the model interprets and distinguishes between question tokens and retrieved content. Therefore, we define the LEA metric as the following:

\begin{table}[t]
\centering
\scriptsize
\caption{Layer–by–layer rank evolution for different models across two CVEs in 2025}%
\label{table:rank_evolution_combined}
\begin{tabular}{@{}clccc@{}}
\toprule
\multicolumn{2}{c}{} & \textbf{Llama-3.2-3B}%
& \textbf{Mistral-Small-24B}%
& \textbf{Gemma-3-27b} \\ \midrule
\multicolumn{2}{c}{} & \textbf{32 Layers} & \textbf{40 Layers} & \textbf{62 Layers} \\ \midrule

\multirow{9}{*}{\rotatebox{90}{\textbf{CVE-2025-30066}}} 
& & \textit{\textbf{Layer 0}} & \textit{\textbf{Layer 0}} & \textit{\textbf{Layer 0}}\\
& rank $x\theta y$ & 48\% & 48\% &  47\% \\
& rank $xy$      &  50\% & 57\% &  64\% \\ 
\cmidrule(l){2-5}
& & \textit{\textbf{Layers 1-2}} & \textit{\textbf{Layers 1-5}} & \textit{\textbf{Layers 1-23}}\\
& rank $x\theta y$ & 98\% & 95\% &  88\% \\
& rank $xy$      &  98\% & 98\% &  94\% \\ 
\cmidrule(l){2-5}
& & \textit{\textbf{Layers 3-27}} & \textit{\textbf{Layers 6-39}} & \textit{\textbf{Layers 24-61}} \\
& rank $x{\theta}y$ & 100\%  & 100\% & 100\% \\
& rank $xy$ & 100\% & 100\%  & 100\%  \\ \midrule \midrule

\multirow{9}{*}{\rotatebox{90}{\textbf{CVE-2025-24472}}} 
& & \textit{\textbf{Layer 0}} & \textit{\textbf{Layer 0}} & \textit{\textbf{Layer 0}}\\
& rank $x\theta y$ & 54\% & 48\% &  45\% \\
& rank $xy$      &  67\% & 68\% &  65\% \\ 
\cmidrule(l){2-5}
& & \textit{\textbf{Layers 1-3}} & \textit{\textbf{Layers 1-5}} & \textit{\textbf{Layers 1-28}}\\
& rank $x\theta y$ & 97\% & 96\% &  86\% \\
& rank $xy$      &  99\% & 98\% &  94\% \\ 
\cmidrule(l){2-5}
& & \textit{\textbf{Layers 4-27}} & \textit{\textbf{Layers 6-39}} & \textit{\textbf{Layers 29-61}} \\
& rank $x{\theta}y$ & 100\% & 100\% & 100\% \\
& rank $xy$ & 100\% & 100\% & 100\%  \\ 
\bottomrule
\end{tabular}
\end{table}

\begin{figure*}[t]
\centering
\includegraphics[scale=0.28]{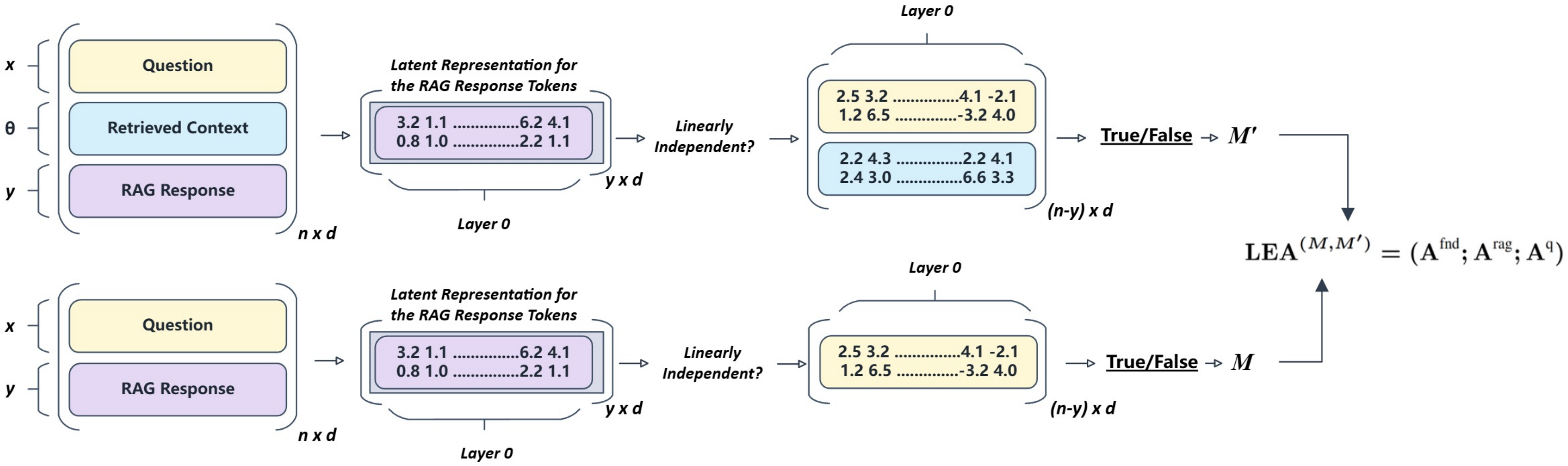}
\caption{The process to compute LEA with the dependent vectors with (top) and without (bottom) retrieved context.}
\label{fig:dependency_process}
\end{figure*}

\newcommand{\Sset}{\mathcal{S}}
\newcommand{\fnd}{\text{fnd}}
\newcommand{\rag}{\text{rag}}
\newcommand{\qq}{\text{q}}

\begin{equation}
    A^{(M,M')} = \frac{1}{L_y} \sum_{i=1}^{L_y} 
    \mathbf{1}\!\left[\, \delta_i^{M} \;\land\; \delta_i^{M'} \,\right]
\end{equation}

\begin{equation}
    LEA^{(M,M')} = (A^{(1,1)}; A^{(1,0)}; A^{(0,0)})
    = (A^{\fnd}; A^{\rag}; A^{\qq})
    \label{equ: lea}
\end{equation}

\noindent Here, $M\!\in\!\{xy\}$ and $M'\!\in\!\{x\theta y\}$ denote distinct sequence configurations, where $M$ refers to sequences without RAG and $M'$ refers to those with RAG. For token $i$ evaluated in sequence $M$ and $M'$, $\delta_i^{M}\in\{0,1\}$ and $\delta_i^{M'}\in\{0,1\}$ equals $1$ iff appending that token’s vector representation
to the question matrix $Q$ increases its rank; otherwise it is $0$.
$L_y$ denotes the number of tokens in the generated response sequence $y$
(i.e., $L_y = |y|$). Figure \ref{fig:dependency_process} shows how we quantify the dependency of the generated tokens in our proposed metric. If a token initially marked as independent (i.e., True) that becomes dependent (i.e., False) after adding $\theta$, this indicates that the retrieved context introduced a new dependency: $A^{\rag} = A^{(1,0)}$. In this case, $\theta$ contributes meaningful contextual information influencing that token. Conversely, if a token remains independent both before and after adding $\theta$, it suggests that the token's interpretation is self-contained within the original question and unaffected by the retrieved content: $A^{\fnd} = A^{(1,1)} $. If a token remains dependent (i.e., False to False), this implies that the dependence is solely attributed to the information present in the original input $x$, and the retrieved content did not impact this relationship: $A^{\qq} = A^{(0,0)} $. Lastly, if adding $\theta$ changes a token from False (dependent) to True (independent), this should not occur as it indicates inconsistency. More specifically, a token initially identified as dependent should not become independent with the addition of more context. 
The progressive transformation from independent to dependent in the responses at layer-0 serves as a measure in understanding the foundational behavior of LLMs in vulnerability analysis, and hence, the proposed LEA metric (equ. \ref{equ: lea}).

\begin{figure*}[t]
    \centering

        \begin{subfigure}{0.75 \textwidth}
        \centering
        \includegraphics[width=\linewidth]{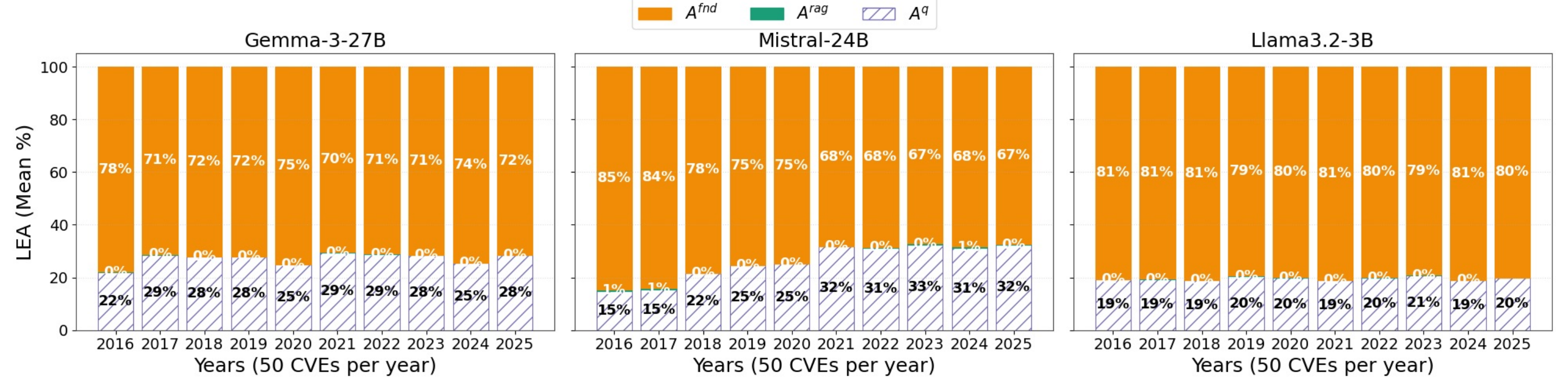}
        \caption{LEA for responses without any retrieved context}
        \label{fig:entire_no_theta}
    \end{subfigure}

    \vspace{0.5em}
    
    \begin{subfigure}{0.75 \textwidth}
        \centering
        \includegraphics[width=\linewidth]{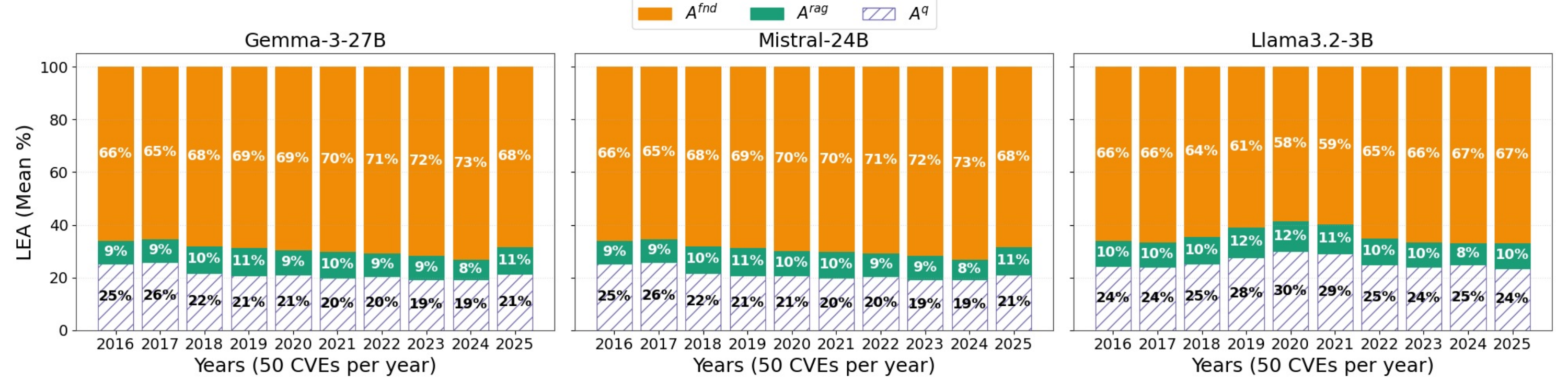}
        \caption{LEA for responses with generic retrieved context}
        \label{fig:entire_generic}
    \end{subfigure}

    \vspace{0.5em}
    
    \begin{subfigure}{0.75 \textwidth}
    \centering
    \includegraphics[width=\linewidth]{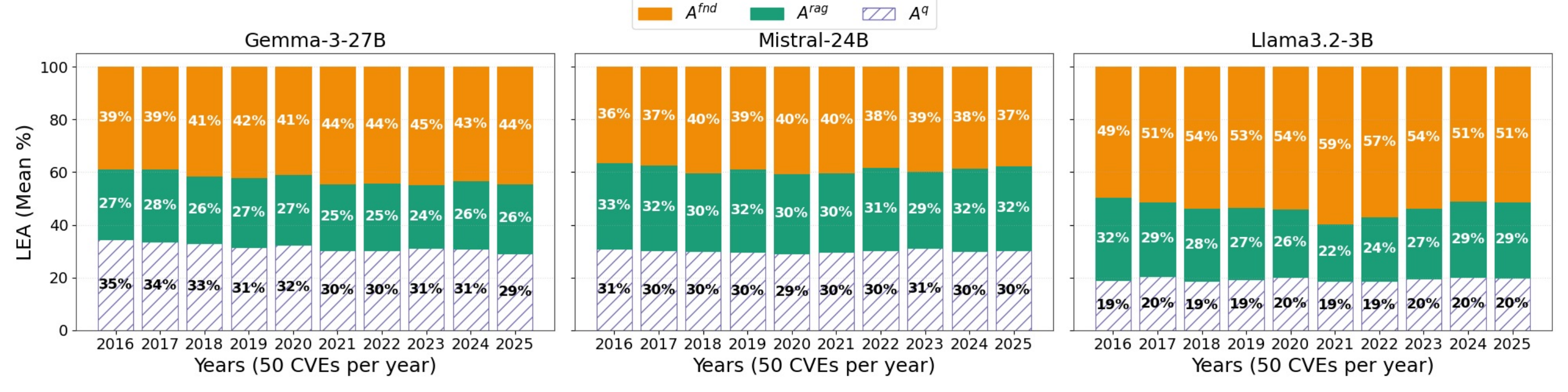}
    \caption{LEA for responses with valid retrieved context}
    \label{fig:entire_ideal}
    \end{subfigure}

    \caption{Comparison of LEA distribution of non-retrieval vs. generic retrieval vs. valid retrieval.}
    \label{fig:lea_entire}
\end{figure*}

\section{Experimental Design}
\label{sec/methodology}

\subsection{Dataset}
To evaluate the effectiveness of the proposed LEA metric for vulnerability analysis, we curated a dataset of 500 high- and critical-severity CVEs reported over the past decade (2016–2025) with a CVSS score above 8.5. To analyze models' performance across different temporal contexts, we selected 50 CVEs per year, allowing us to observe how the models handle the most recent vulnerabilities compared to older ones. For each CVE, we considered three distinct retrieval scenarios:  1) a valid retrieval, 2) a generic retrieval, and 3) an incorrect retrieval. In the valid scenario, we consider the LLM retrieves only the most relevant and verified information from the NVD website \cite{nvd2024}, which is the `Description' segment of each CVE. This serves as a benchmark to evaluate the LEA distribution under optimal retrieval conditions. For the generic retrieval case, we assume that the LLM does not have knowledge of the specific CVE and instead returns generalized information about CVEs. During preliminary research, we observed that a standard web search on ``what a CVE is?'' often returns a general description similar to that provided by RedHat \cite{redhat}. Therefore, we used the following description as the source for the generic RAG scenario:

\begin{promptbox}
    \centering {\it \scriptsize CVE, short for Common Vulnerabilities and Exposures, is a list of publicly disclosed computer security flaws. When someone refers to a CVE, they mean a security flaw that's been assigned a CVE ID number.}
\end{promptbox}

Finally, in the incorrect retrieval scenario, we assume that the LLM returns erroneous or misleading information about a vulnerability. This can occur when the query involves a non-existent CVE ID, causing the model to retrieve information from CVEs that share similar numeric identifiers. For instance, consider a query for CVE-2027-30066, which does not exist. In this case, the model often retrieves CVEs with the same 30066 suffix but from different years. This behavior was observed during our experiments with ChatGPT’s WebSearch tool, where retrieved sources frequently matched the numeric suffix of the queried ID rather than the correct year. Based on this observation, we curated the dataset for the incorrect retrieval scenario to reflect these types of suffix-based mis-retrievals, providing a realistic test of the LEA metric under challenging conditions.

Moreover, since the selected LLMs have a training cut-off up until mid-2024, we expect them to rely more heavily on RAG for the 2025 CVEs. Note that older vulnerabilities are not guaranteed to be consistently represented in the models’ pre-training corpora due to their large volume (over 300,000). Therefore, some years may exhibit a similar LEA distribution of RAG-generated tokens.
It is worth noting that we conducted our experiments on a system equipped with two Intel Xeon E5-2650 CPUs, 256 GB of RAM, two NVIDIA Tesla P40 GPUs, and one NVIDIA Tesla V100 GPU.

\subsection{Experimental Settings}
\label{sec: experimnt_design}

For our experiments, first, we designed a prompt to ask about exploitation of different CVEs, along with the retrieved context, to obtain the generations. We tested on three different LLMs of varying sizes: \textit{Gemma-3-27B-IT}, \textit{Mistral-Small-24B-Instruct-2501}, and \textit{LLaMA-3.2-3B-Instruct}. Note that for the generations, we get the highest probable token (i.e., the token with the highest logit value) to get the most deterministic responses from the LLM. 

We put each CVE-ID in the question, and we will put the retrieved content in the $<<RAG>>$ block. The designed prompt is as follows:

\begin{promptbox}

{\it \footnotesize

\verb|<<Query>>|

You are a cybersecurity expert. How can an attacker exploit CVE-xxx-xxx? Do not be verbose and answer carefully.

\verb|<</Query>>|

\vspace{4pt}

\verb|<<RAG>>|

Retrieved Context

\verb|<</RAG>>|

\vspace{4pt}

\verb|<<Response>>|

}

\end{promptbox}

After getting the generations, we now have the question ($x$), the retrieved context ($\theta$), and the generated response ($y$). 
Our investigation into linear dependence at layer-0 aims to determine whether the model inherently relies on this raw contextual information for subsequent token generation. This layer is critical because it directly processes the input embeddings augmented by positional encodings, which reflects the model's initial view of token identity and order. Therefore, we will use the proposed LEA metric (refer to Equ. \ref{equ: lea}) to get the LEA distributions under different RAG scenarios.

\begin{figure*}[t]
    \centering
    \begin{subfigure}{0.75 \textwidth}
        \centering
        \includegraphics[width=\linewidth]{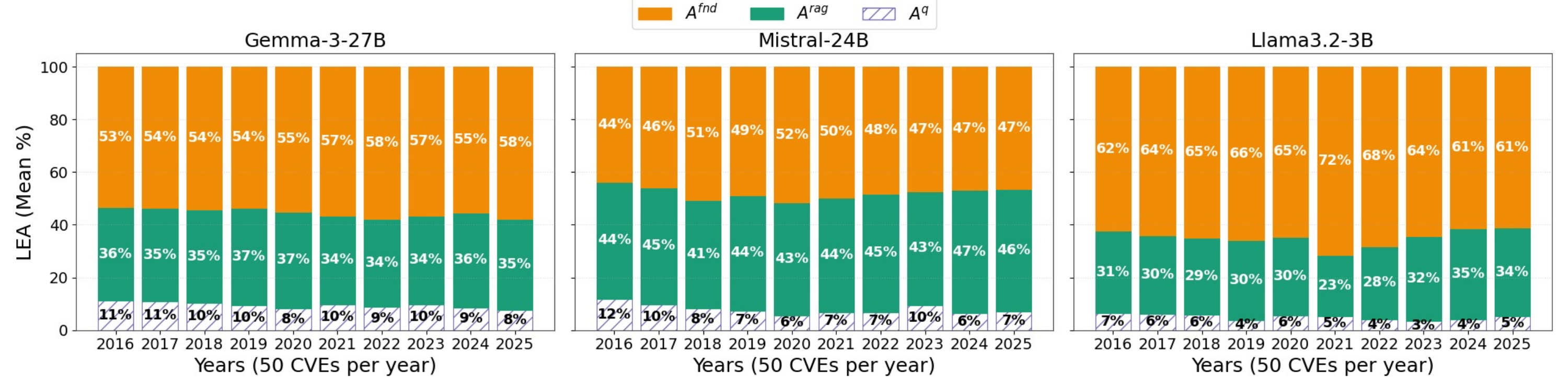}
        \caption{Filtered LEA distribution for valid response with valid RAG}
        \label{fig:results_filtered_ideal_theta_ideal_y}
    \end{subfigure}
    
    \vspace{0.5em}
    
    \begin{subfigure}{0.75 \textwidth}
        \centering
        \includegraphics[width=\linewidth]{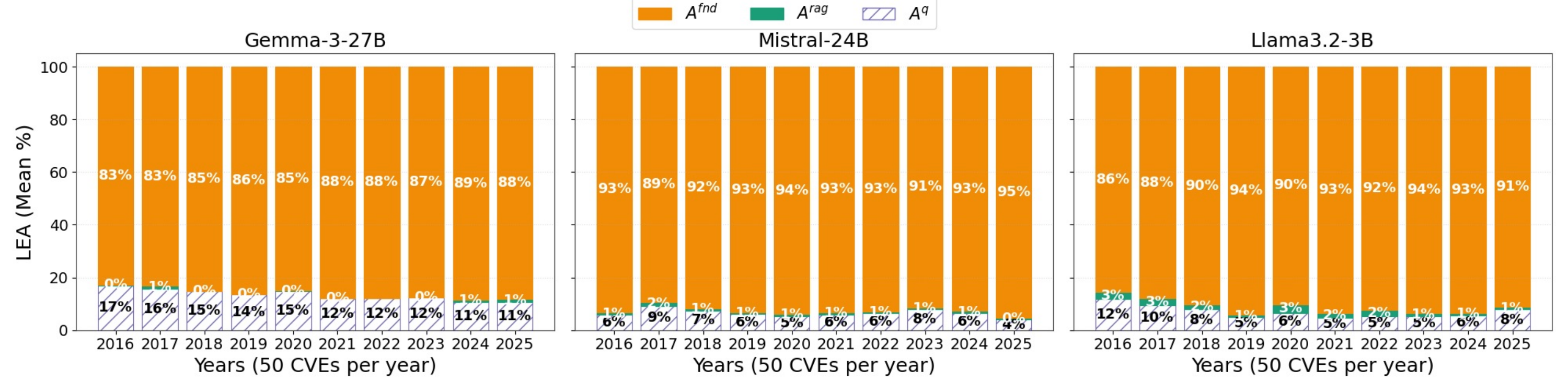}
        \caption{Filtered LEA distribution for valid response with generic RAG}
        \label{fig:results_filtered_generic_theta_ideal_y}
    \end{subfigure}
    
    \vspace{0.5em}
    
    \begin{subfigure}{0.75 \textwidth}
        \centering
        \includegraphics[width=\linewidth]{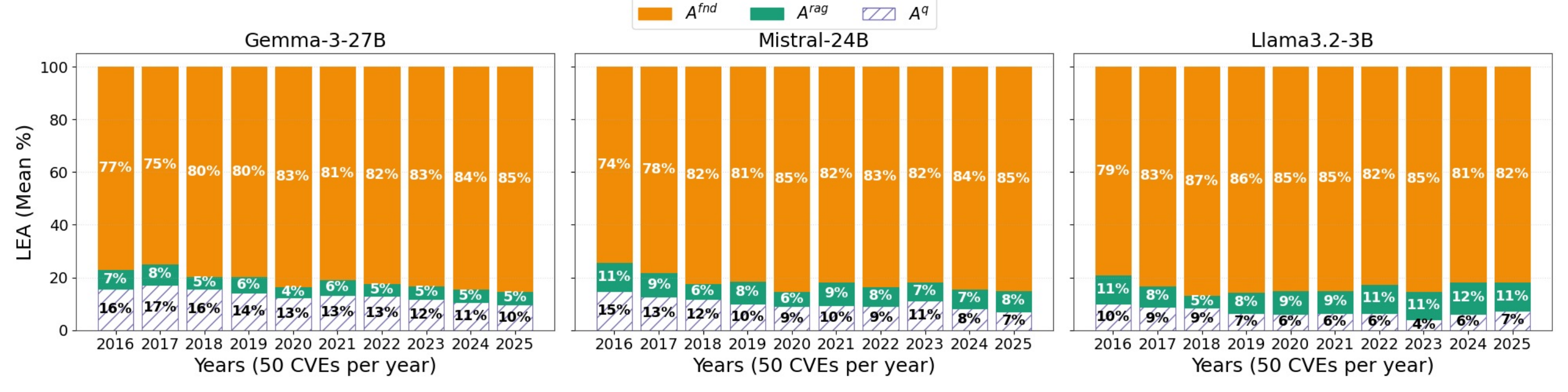}
        \caption{Filtered LEA distribution for valid response with incorrect RAG}
        \label{fig:results_filtered_wrong_theta_ideal_y}
    \end{subfigure}
    
    \caption{Comparison of filtered LEA distribution of the valid response with different RAG context.}
    \label{fig:filtered_multiple_theta_ideal_y}
\end{figure*}

\section{Usage of LEA for CVE Analysis}
\label{sec/results}

To evaluate LEA's utility and robustness, we present four complementary sets of results. First, we will get the LEA distribution over the entire response and compare the non-retrieval vs. generic retrieval vs. valid retrieval. Next, we apply filtering to focus on content-bearing tokens and compare the LEA distribution to see whether the response match to the right retrieval. Third, we compute LEA’s expected distribution over generations when optimal retrieval is known. Fourth, we compare the case of valid retrieval vs. incorrect retrieval and raise concerns about the blind use of retrievals for vulnerability analysis to security analysts. Finally, we discuss the usability of LEA for SOC workflows.

\begin{figure*}[t]
    \centering
    \begin{subfigure}{0.75 \textwidth}
        \centering
        \includegraphics[width=\linewidth]{./figures/results_filtered_ideal_theta_ideal_y.pdf}
        \caption{Filtered LEA distribution for valid RAG with valid response}
        \label{fig:results_filtered_ideal_theta_ideal_y_2}
    \end{subfigure}
    
    \vspace{0.5em} 
    
    \begin{subfigure}{0.75 \textwidth}
        \centering
        \includegraphics[width=\linewidth]{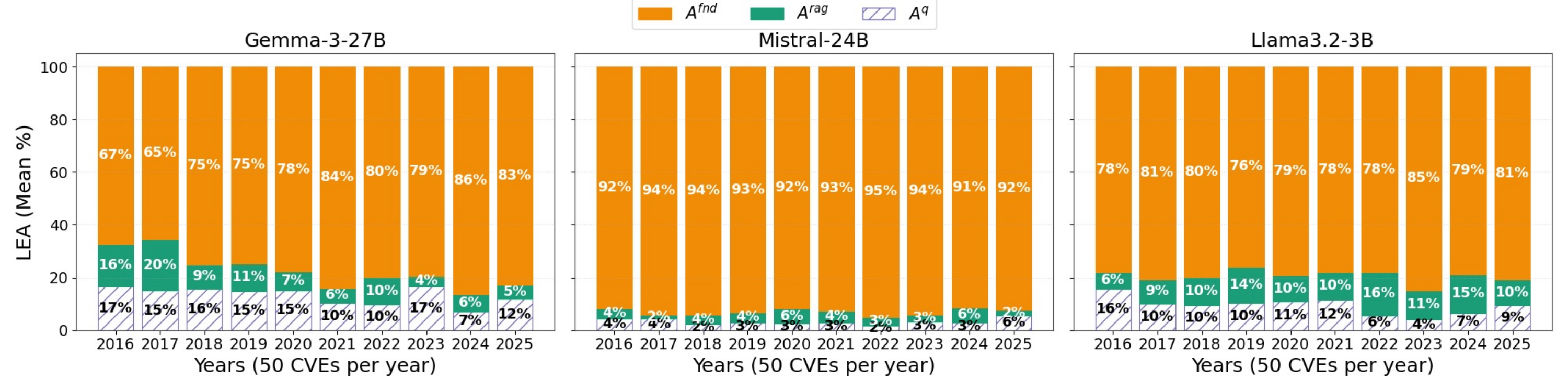}
        \caption{Filtered LEA distribution for valid RAG with generic response}
        \label{fig:results_filtered_ideal_theta_generic_y}
    \end{subfigure}
    
    \vspace{0.5em}
    
    \begin{subfigure}{0.75 \textwidth}
        \centering
        \includegraphics[width=\linewidth]{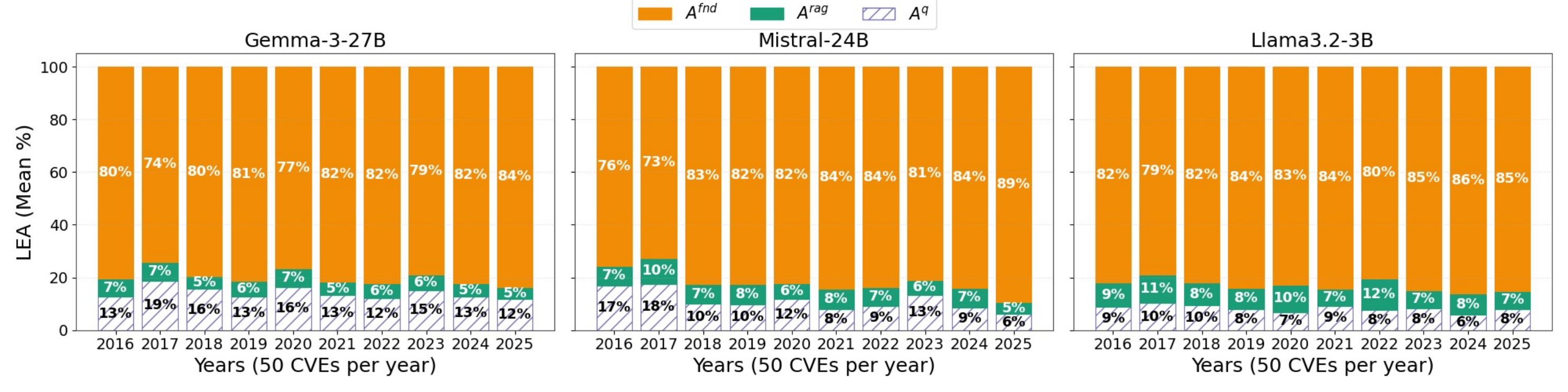}
        \caption{Filtered LEA distribution for valid RAG with incorrect response}
        \label{fig:results_filtered_ideal_theta_wrong_y}
    \end{subfigure}
    
    \caption{Comparison of filtered LEA distribution of the valid ground-truth RAG with different response generations.}
    \label{fig:filtered_ideal_rag_different_y}
\end{figure*}

\subsection{Can LEA perform under different retrieval and non-retrieval scenarios?}

In this section, we compute the LEA distributions for three different cases: 1) baseline: the response without any retrieved context, 2) generic retrieval: the response with a generic RAG describing what a CVE is, and 3) valid retrieval: the response with the most relevant NVD description across CVEs from the past 10 years. Figure \ref{fig:lea_entire} summarizes mean LEA scores across 50 CVEs per year for multiple LLMs, to examine performance changes before and after knowledge cutoffs (Aug 2024 for Gemma3-27B and Oct 2023 for Mistral-24B).  

From the results, we can see that the LEA distribution for the responses without any retrieved context show a significant dependence on each model's internal knowledge, which is expected as there is no retrieval, and about 15\% to 30\% on the input query tokens. For the generic retrieval case, we see some level of dependence on the RAG tokens ($A^{\rag}$), but still a significant distribution over the model's internal knowledge, showing that generic retrieval does not really attribute to the model's generation. For the valid retrieval case, the results reveal a broadly distributed dependency of tokens across all the years and the models. This shows that the models lean more heavily on retrieved information when provided with valid, informative content. From the results, we can also see that the smaller model exhibits a greater reliance on its \textit{LEA}($A^{\fnd}$) compared to the larger models. This suggests that, despite its limited capacity, the smaller model tends to utilize pre-trained knowledge more than the retrieved content. 

Note that in the experiments, the query provides only the CVE-ID (e.g., CVE-2025-30066) without additional context. Given the vast CVE space (300K+ IDs), it is improbable for any LLM to memorize detailed knowledge for each. This explains the relatively uniform dependency across years and underscores the importance of external retrieval for accurate, context-aware responses in cyber operations.
These values reveal that when valid retrieval is happening, the expected distribution for a verified response is in Figure \ref{fig:entire_ideal}, and demonstrate the effectiveness of LEA in showcasing the differences under these scenarios. In Sec \ref{sec: discussion}, we demonstrate the validity of LEA in differentiating these across all 500 CVEs. Next, we evaluate whether LEA matches the right retrieval when exposed with different RAG content.

 \begin{table*}[t]
    \centering
    \scriptsize
    \caption{LEA's distribution for a CVE over different generations with the NVD description as the optimal retrieval.}
    \label{tab:specific_examples}

    \renewcommand{\arraystretch}{1.0}  
    \setlength{\tabcolsep}{6pt}        

    \begin{tabular}{@{} c p{0.8\linewidth} @{}}
        \toprule
        \textbf{CVE-2024-5756} &  \\
        \toprule

         NVD-Description  & {The Email Subscribers by Icegram Express – Email Marketing, Newsletters, Automation for WordPress \& WooCommerce plugin for WordPress is vulnerable to time-based SQL Injection via the db parameter in all versions up to, and including, 5.7.23 due to insufficient escaping on the user supplied parameter and lack of sufficient preparation on the existing SQL query. This makes it possible for unauthenticated attackers to append additional SQL queries into already existing queries that can be used to extract sensitive information from the database.} \\ 
         \midrule

         \textbf{\textit{Gemma-3-27B} Output} & {\footnotesize \textbf{$LEA =(A^{\fnd}; A^{\rag}; A^{\qq} )$}} \\

         \midrule
         \textit{Non-Retrieval-Response} & {CVE-2024-5756 is a remote code execution vulnerability in VMware ESXi. An unauthenticated attacker with network access can exploit it by sending a crafted request to the OpenSLP service. This allows them to execute arbitrary code on the ESXi host. The vulnerability is due to a heap overflow in the handling of SLP messages. \colorbox{green!25}{\it \textbf{LEA = (73\%; 18\%; 9\%)}}} \\
         \midrule
         \textit{Generic-RAG-Response}  & {CVE-2024-5756 is a remote code execution vulnerability in VMware ESXi. An unauthenticated attacker with network access can exploit it by sending a crafted request to the OpenSLP service. Successful exploitation allows arbitrary code execution on the ESXi host. Mitigation involves disabling or removing the OpenSLP service. \colorbox{green!25}{\it \textbf{LEA = (83\%; 17\%; 0\%)}}} \\
         \midrule
         \textit{Incorrect-RAG-Response}  & {An attacker can exploit CVE-2024-5756 by crafting a malicious link or webpage. If a WordPress administrator clicks this link while logged in, the attacker can trigger unauthorized AJAX actions within the Digital Publications plugin due to the missing nonce validation. This could allow the attacker to perform actions with the administrator's privileges. \colorbox{green!25}{\it \textbf{LEA = (73\%; 18\%; 9\%)}}} \\
         \midrule
         \textit{Valid-RAG-Response}  & {An attacker can exploit CVE-2024-5756 by injecting malicious SQL code through the `db` parameter. This allows them to extract sensitive information from the WordPress database without authentication, due to insufficient input sanitization and query preparation in the Email Subscribers by Icegram Express plugin (<= 5.7.23). A time-based SQL injection technique can be used to infer data. \colorbox{green!25}{\it \textbf{LEA = (35\%; 55\%; 10\%)}}} \\
        \bottomrule
    \end{tabular}
\end{table*}

\subsection{Does LEA match to the right retrieval?} 

To validate LEA's ability in matching to the right retrieval, we start by comparing the generated response from the valid retrieval with different RAG content, namely valid, generic, and incorrect. In the previous section, we demonstrated the LEA distributions over different retrieval scenarios on the entire generated tokens. However, tokens such as ``on'' or ``the'' contribute little semantic value, despite influencing the overall distribution. Therefore, to sharpen our analysis, we focus on semantically meaningful content. We apply stop-word filtering (e.g., removing tokens such as ``of'', ``the'') and introduce a probability-based thresholding mechanism.

\begin{table}[H]
\centering
\scriptsize
\caption{Token-level probability deltas ($\Delta p = x\theta y - xy$)}
\begin{tabular}{@{}ccccc@{}}
\toprule
\textbf{Response ID} & \textbf{Token} & $x\theta y$ & $xy$ & $\Delta p$ ($x\theta y - xy$) \\
\midrule
257   & \texttt{attacker} & 1.000 & 1.000 & 0.000 \\
258   & \texttt{exploits} & 0.770 & 0.001 & 0.769 \\
259   & \texttt{CVE}      & 1.000 & 1.000 & 0.000 \\
260   & \texttt{-}        & 1.000 & 1.000 &  0.000 \\
261–264 & \texttt{2025}   & 1.000 (each) & 1.000 (each) & 0.000 (each) \\
\bottomrule
\end{tabular}
\label{tab:token_deltas}
\end{table}

For threshold filtering, we show the token-level probability deltas ($\Delta p = x\theta y - xy$) in Table \ref{tab:token_deltas}, where $x\theta y$ and $xy$ represent token probabilities of the generated response before and after removing the RAG context, respectively. These probabilities are taken from the final model layer (post-Softmax). The tokens that produce $\Delta p \leq 0$ are filtered out, as they contribute negligible discriminative information in our evaluation settings. This dual filtering ensures that our metric emphasizes content-containing tokens that contribute meaningfully to the model’s output, thereby reducing noise and yielding a more accurate assessment of retrieval quality.

By comparing Figure \ref{fig:entire_ideal} with Figure \ref{fig:results_filtered_ideal_theta_ideal_y}, we can see that filtering out non-informative tokens results in a noticeable decrease in the proportion of $A^{\qq}$. This outcome is expected, as the question tokens primarily consist of the CVE identifier, which remain invariant with or without RAG context. While these identifiers are syntactically important, they carry minimal semantic weight in guiding generation. From a practical standpoint, security analysts are less concerned with surface-level identifiers and more interested in whether and how the retrieved context influences the model’s substantive understanding of the vulnerability.
This shift reinforces the value of our filtered metric: it demonstrates the extent to which informative content (rather than syntactic tokens) relies on external retrieval. Moreover, by analyzing tokens with significant $\Delta p$ values, such as `exploits' in Table \ref{tab:token_deltas}, we can identify which parts of the model’s output are context-dependent versus internally derived (refer to Figure \ref{fig:example}).

Now, to assess LEA's validity in matching to the right retrieval, we start by comparing our valid response with multiple RAG context, namely valid, generic, and incorrect. Figure \ref{fig:filtered_multiple_theta_ideal_y} shows the results. From these results, we observe a significant decrease in $A^{\rag}$ when comparing the valid response tokens with the generic and incorrect retrieval. This supports the validity of our approach by demonstrating that the proposed LEA metric is sensitive to the informativeness of the retrieved context. In other words, when meaningful retrieval is absent, the model exhibits a measurable drop in $A^{\rag}$, reinforcing LEA’s utility in distinguishing context-driven generation from reliance on internal knowledge.

 Overall, this demonstrates the effectiveness of the proposed LEA metric. The strongest dependence of $A^{\rag}$ emerges when the valid response is compared against the valid retrieval, which demonstrates its ability to correctly align with the most relevant source. Building on this, next, we consider the scenario where the ground-truth retrieval is available, and examine how the LEA distribution should ideally behave when dealt with the most verified generated responses.

\begin{figure*}[t]
\centering
\includegraphics[scale=0.3]{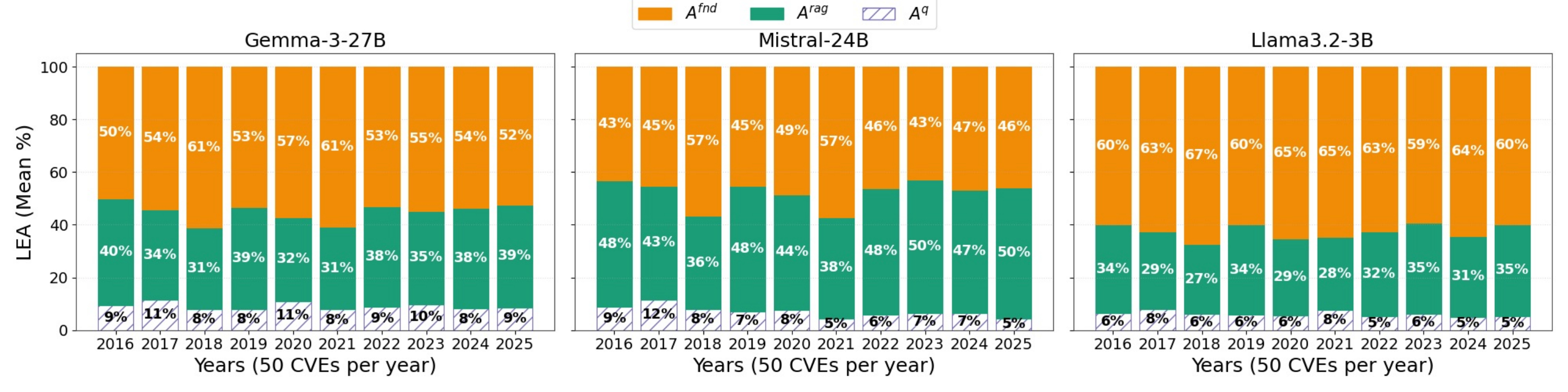}
\caption{LEA distribution for the incorrect retrieval of a CVE in the query}
\label{fig:results_wrong_wrong}
\end{figure*}

\begin{figure*}[t]
    \centering
    \begin{subfigure}{0.32 \textwidth}
        \centering
        \includegraphics[width=\linewidth]{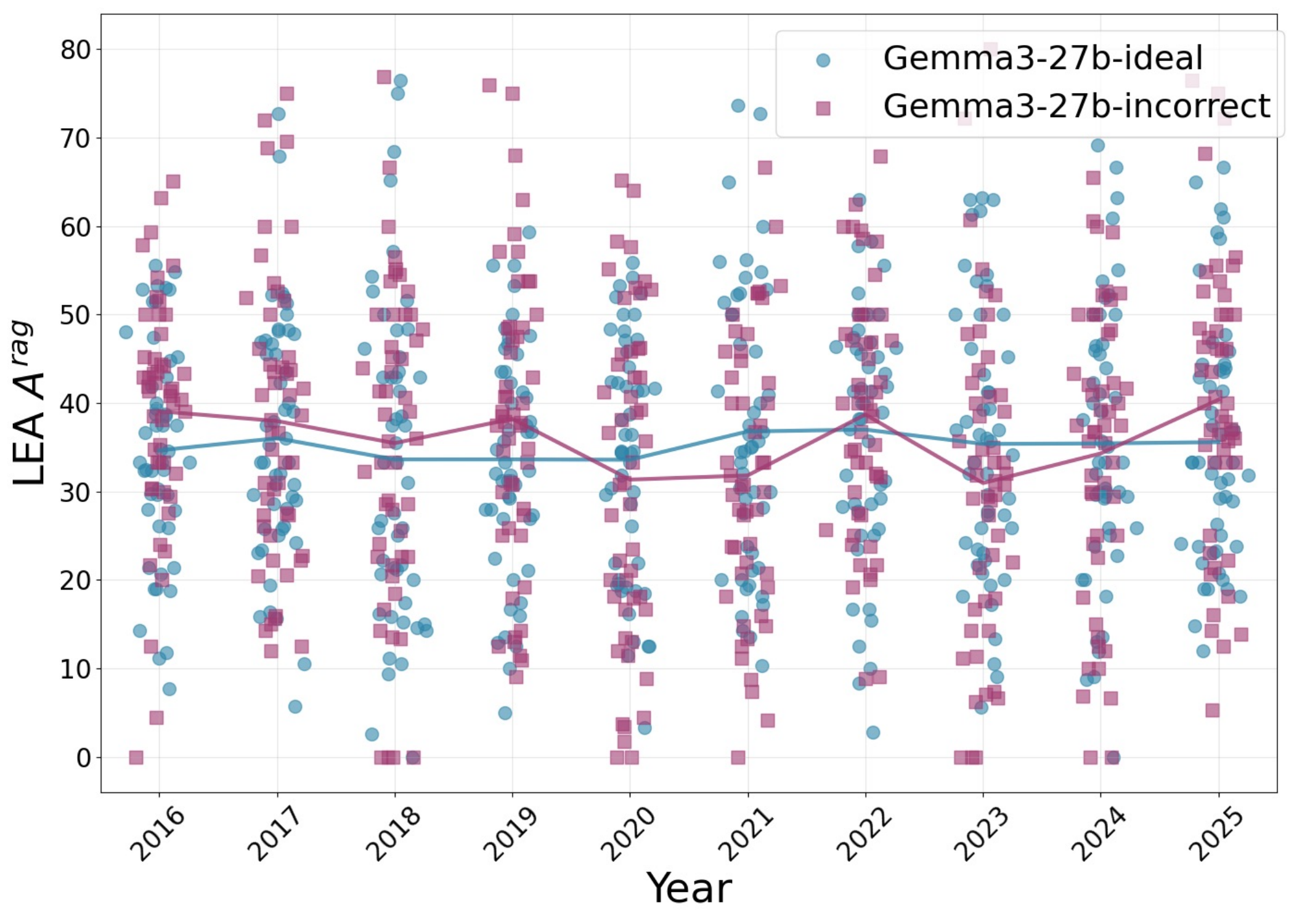}
        \caption{Gemma3-27b's $A^{\rag}$ distribution}
        \label{fig:scatter_gemma_ideal_wrong}
    \end{subfigure}
    \hfill
    \begin{subfigure}{0.32 \textwidth}
        \centering
        \includegraphics[width=\linewidth]{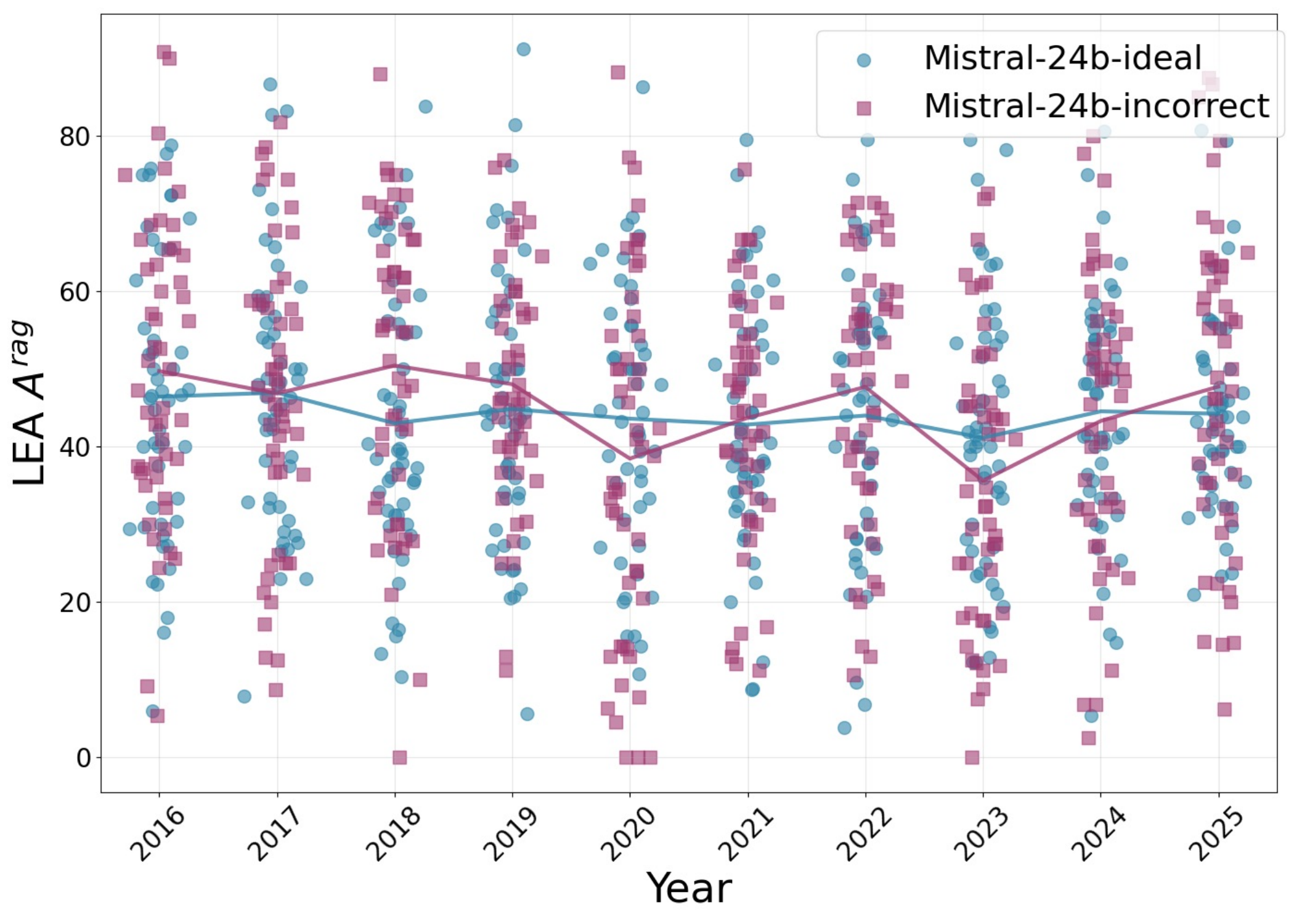}
        \caption{Mistral-24b's $A^{\rag}$ distribution}
        \label{fig:scatter_mistral_ideal_wrong}
    \end{subfigure}
    \hfill
    \begin{subfigure}{0.32 \textwidth}
        \centering
        \includegraphics[width=\linewidth]{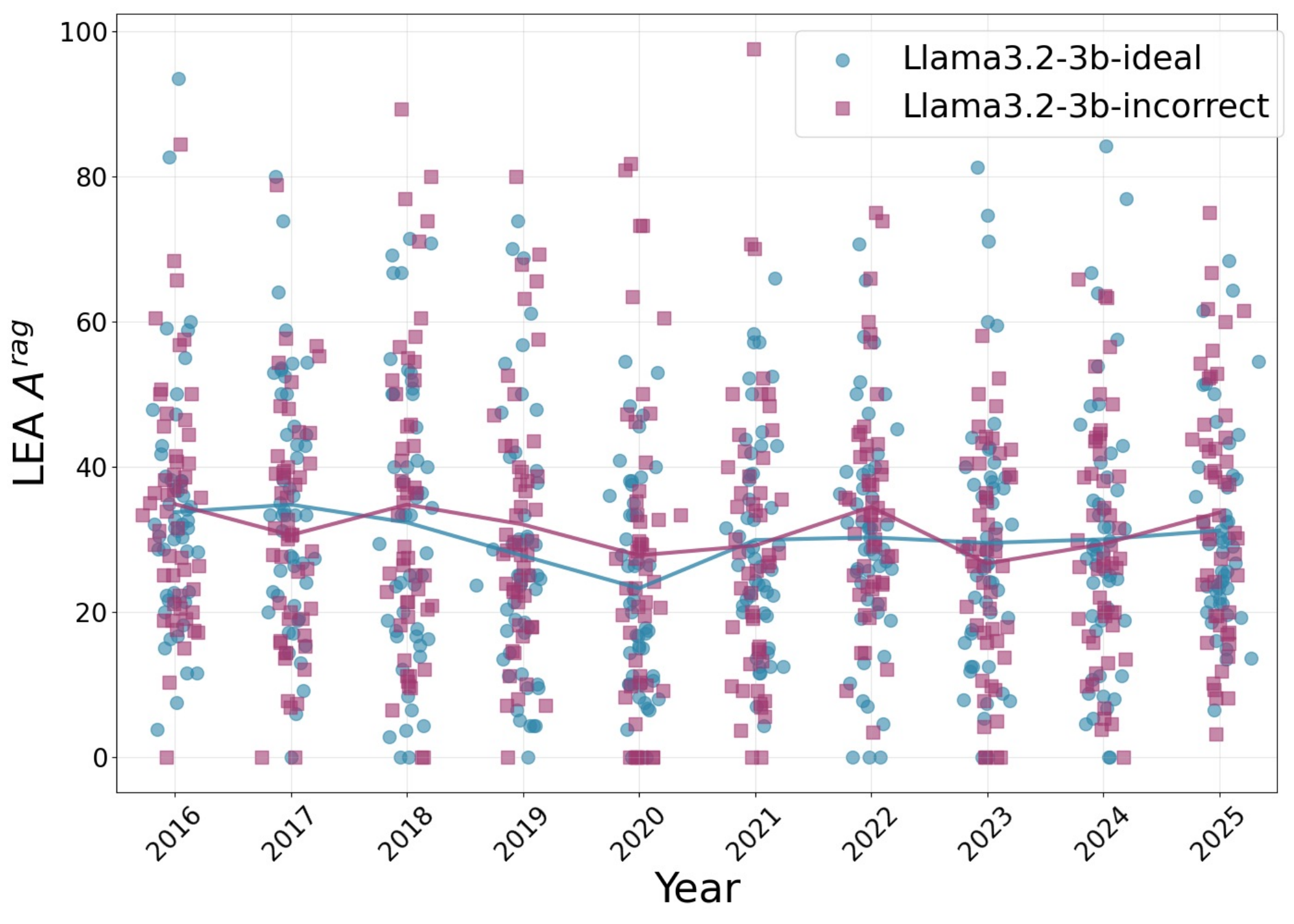}
        \caption{Llama3.2-3b's $A^{\rag}$ distribution}
        \label{fig:scatter_llama_ideal_wrong}
    \end{subfigure}
    
    \caption{Comparison of LEA's $A^{\rag}$ distribution of valid-retrieval-response vs. incorrect-retrieval-response over the years.}
    \label{fig:results_scatter_ideal_wrong}
\end{figure*}

\subsection{Can LEA identify bad generations given the optimal retrieval?} 

So far, our analysis has examined LEA's ability differentiate different retrieval scenarios and to align with the correct retrieval. In this section, we examine the LEA distribution for the various retrieval scenarios when the ground-truth / optimal retrieval ($\theta$) is known. By assuming access to the optimal, verified retrieval, we can establish a benchmark for how LEA's distribution should ideally look when generating the most accurate response. To do so, we consider the valid retrieved content as the `optimal' retrieval, and we get the LEA's distribution by comparing this content (i.e., NVD description) against multiple generation under different scenarios. Figure \ref{fig:filtered_ideal_rag_different_y} shows the distributions of the verified response (\ref{fig:results_filtered_ideal_theta_ideal_y_2}) vs. the generic response (\ref{fig:results_filtered_ideal_theta_generic_y}) vs. the incorrect response (\ref{fig:results_filtered_ideal_theta_wrong_y}). The comparison clearly illustrates how LEA assigns higher attributions to tokens that are semantically aligned with the optimal retrieval and lower values when the generated content deviates from the verified context.
 We can also see that in the $A^{\rag}$ significantly decreases when using the response that is not following the retrieval content of the valid retrieval (NVD description). This analysis serves as a proxy for what an insightful, contextually relevant response should look like, and highlights how the model internally reasons about a vulnerability when the LEA"s response distribution is not matching to the valid retrieval.  
 
In Table \ref{tab:specific_examples}, we demonstrate how the responses' LEA distribution vary when different generations are involved. As shown, when the response is entirely ignoring the verified NVD embeddings, it suggests that the model either hallucinated or produced a response with minimal factual relevance, which indicates the necessity of verified retrieval for reliable and informative outputs in cybersecurity applications.   
Furthermore, we observe frequent hallucinations or lack of factual detail of responses when valid retrieval is not present. This is expected, LLMs cannot reasonably memorize or recall over 300K CVEs, especially when prompted with only an identifier. Unlike search engines, they lack indexed recall, highlighting the limits of relying solely on internal knowledge. This allows analysts to assess whether a model is synthesizing information based on learned patterns or simply echoing retrievals.

Based on these findings, we draw two conclusions: (1) retrieval is indispensable for producing informative, trustworthy outputs in vulnerability analysis, and (2) the proposed LEA metric offers a robust measure for quantifying model-context reliance, which allows for informed decision-making when interpreting responses about unseen or complex vulnerabilities. This analysis provides a reference point for evaluating the quality of various responses, by quantifying the deviations from the valid behavior (i.e., with optimal retrieval). Next, we assess whether LLMs can detect and recognize individual CVEs when exposed with incorrect retrieval content.

\begin{figure*}[t]
    \centering
    \begin{subfigure}{0.32 \textwidth}
        \centering
        \includegraphics[width=\linewidth]{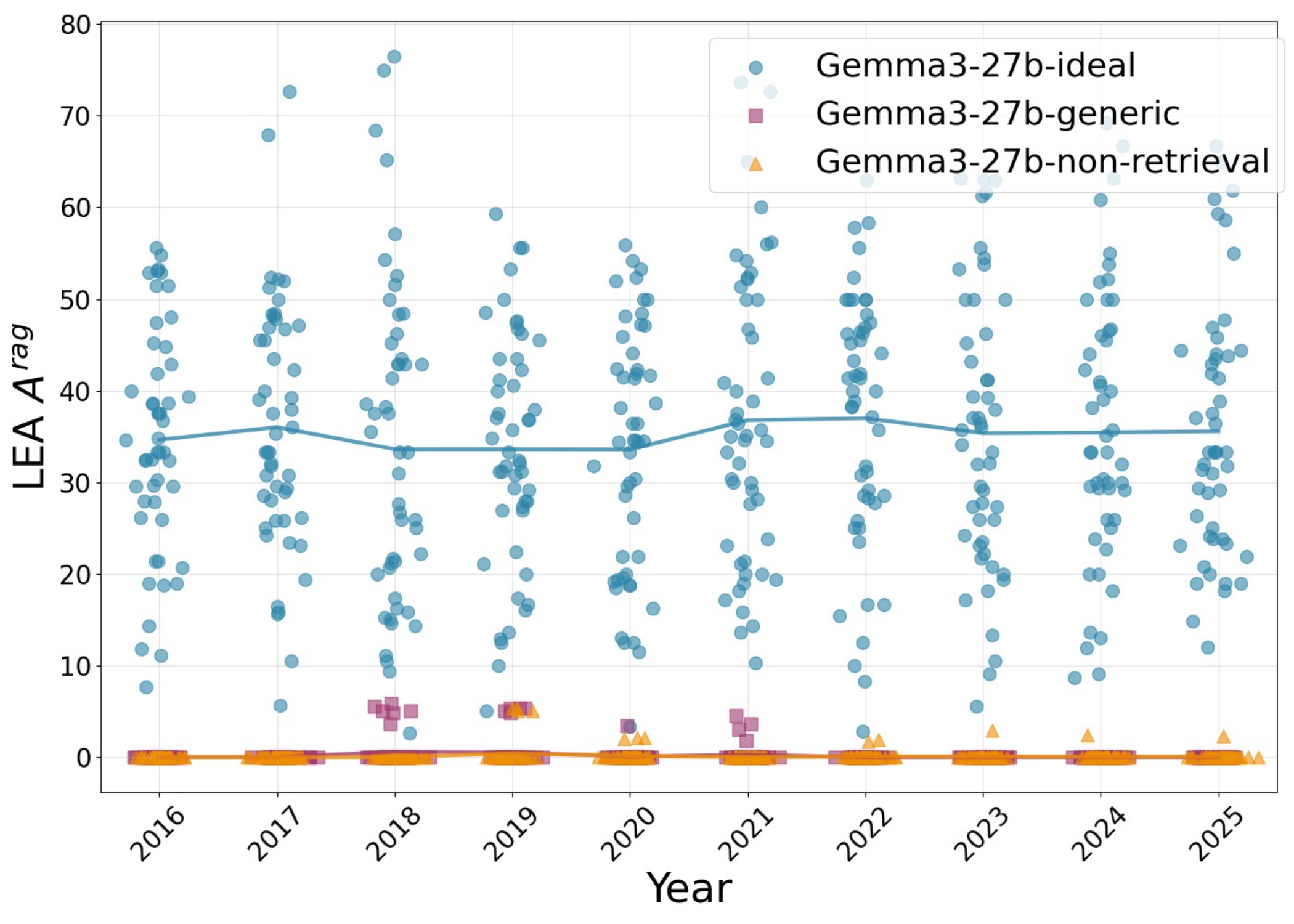}
        \caption{Gemma3-27b's $A^{\rag}$ distribution}
        \label{fig:scatter_gemma_yearly_filtered}
    \end{subfigure}
    \hfill
    \begin{subfigure}{0.32 \textwidth}
        \centering
        \includegraphics[width=\linewidth]{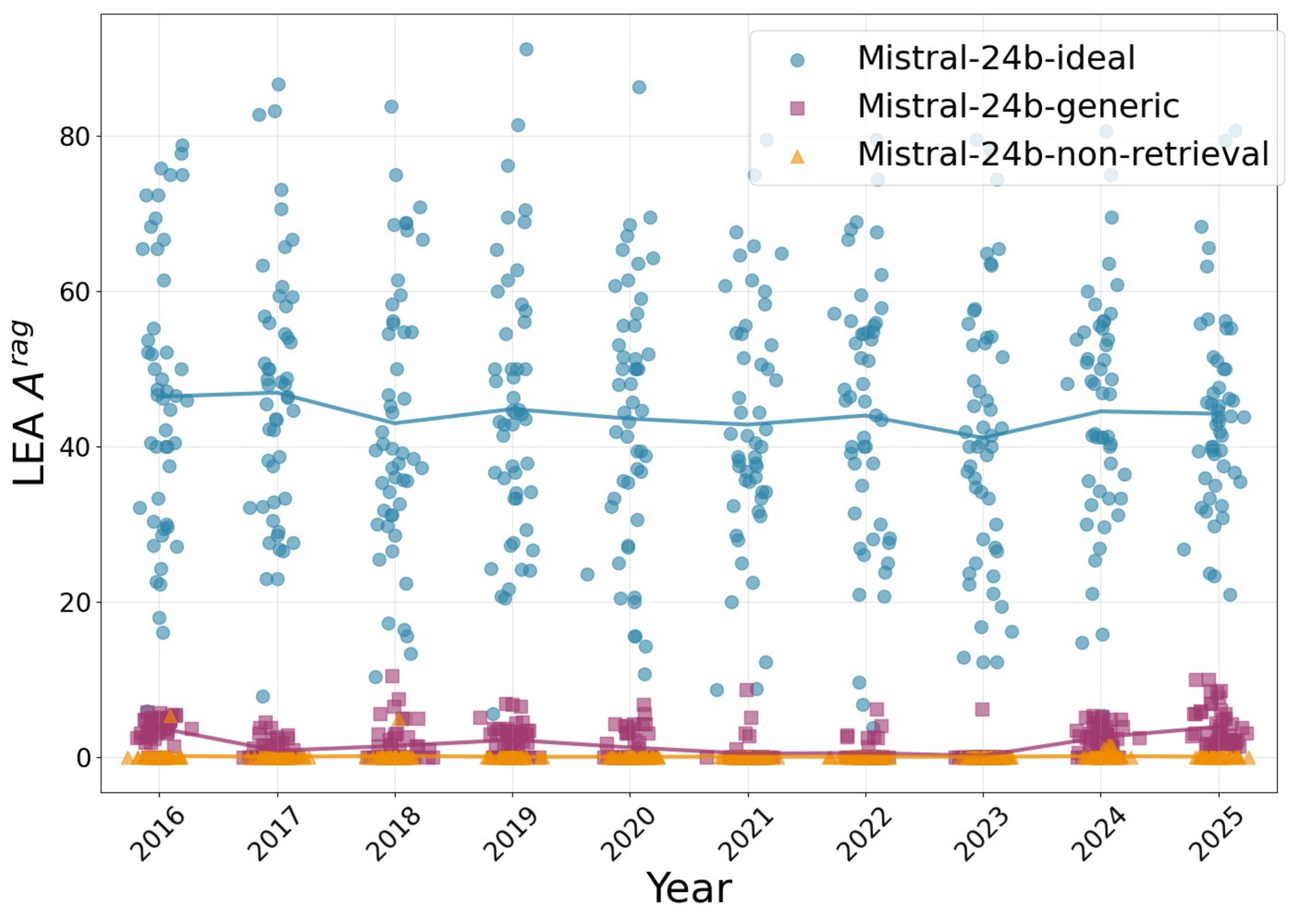}
        \caption{Mistral-24b's $A^{\rag}$ distribution}
        \label{fig:scatter_mistral_yearly_filtered}
    \end{subfigure}
    \hfill
    \begin{subfigure}{0.32 \textwidth}
        \centering
        \includegraphics[width=\linewidth]{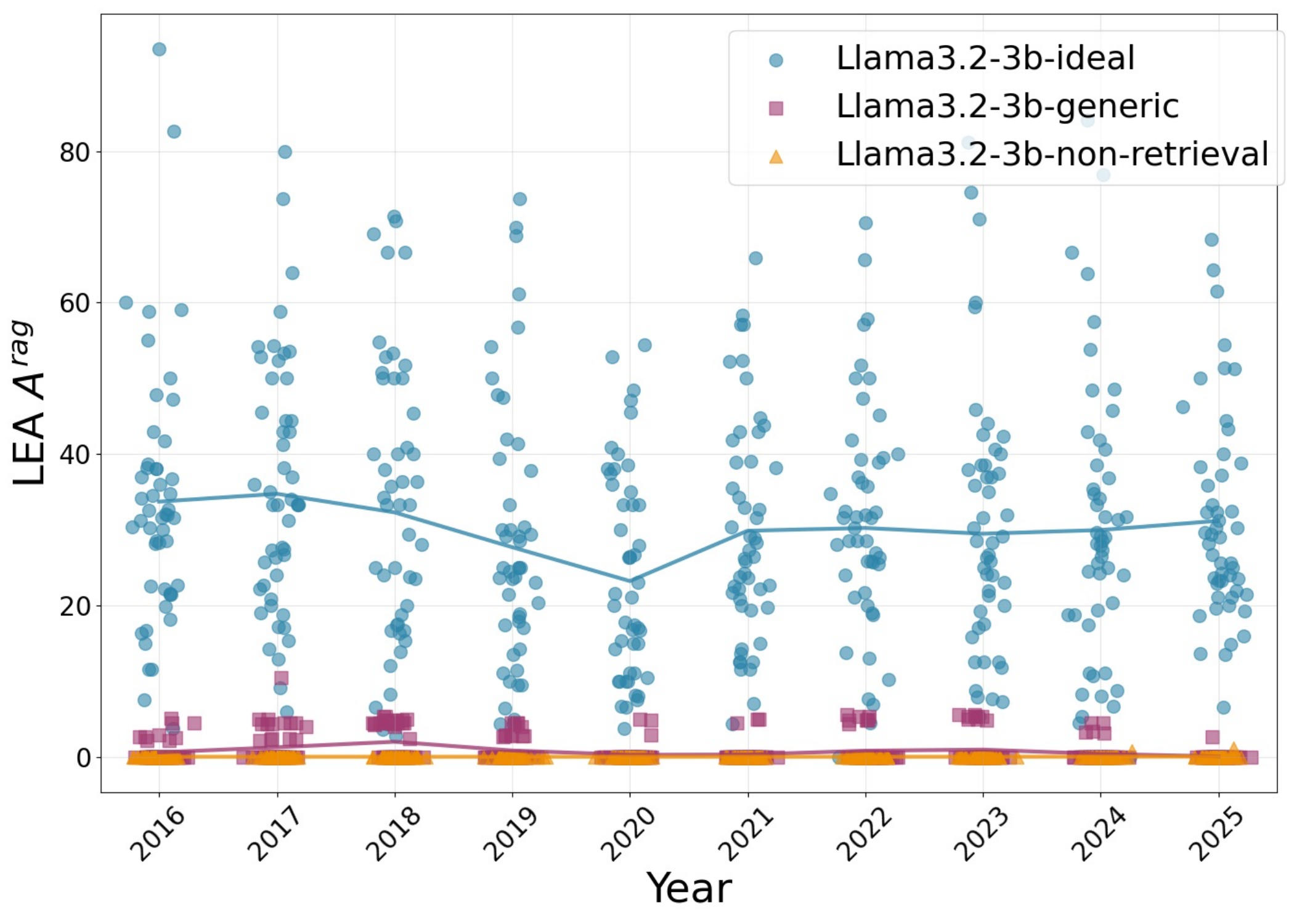}
        \caption{Llama3.2-3b's $A^{\rag}$ distribution}
        \label{fig:scatter_llama_yearly_filtered}
    \end{subfigure}
    
    \caption{Comparison of LEA's $A^{\rag}$ distribution of non-retrieval vs. generic retrieval vs. valid retrieval of all the curated CVEs.}
    \label{fig:results_scatter_lea}
\end{figure*}

\begin{table*}[t]
\centering
\footnotesize
\caption{LEA's Valid vs. Generic/No Retrieval Classification Performance Across Models}
\label{tab:lea_model_comparison}
\begin{tabular}{lcccccc}
\hline
\textbf{Model} & \textbf{Threshold (\%)} & \textbf{Train Acc.} & \textbf{Train F1} & \textbf{Test Acc.} & \textbf{Test F1} & \textbf{ROC AUC} \\ \hline
Gemma3-27b     & 12.50 & 0.950 & 0.925 & 0.957 & 0.933 & 0.979 \\
Mistral-24b    & 13.30 & 0.974 & 0.961 & 0.967 & 0.948 & 0.983 \\
Llama3.2-3b    & 12.30 & 0.919 & 0.879 & 0.927 & 0.892 & 0.952 \\ \hline
\end{tabular}
\end{table*}

\subsection{Do LLMs recognize individual CVEs? A rising concern for security analysts}

We demonstrated LEA's effectiveness in modeling the expected distribution of when the best retrieved information is present. This raises a critical question: given a pool of over 300,000 CVEs, do LLMs actually recognize individual CVEs, or do they simply rely on the context provided, even if it is incorrect? To answer this, we will ask about the 500 CVEs we curated, but this time, we will feed the LLM with information about another CVE (i.e., incorrect retrieval) to see whether the model blindly relies on the retrieved context. 

Figure \ref{fig:results_wrong_wrong} illustrates the LEA distribution under these conditions. Despite being provided with incorrect retrievals, the distribution closely mirrors that of the valid case (see Figure \ref{fig:results_filtered_ideal_theta_ideal_y}). Further confirmation comes from Figure \ref{fig:results_scatter_ideal_wrong}, which shows a strong overlap between the valid and incorrect distributions across all 500 CVEs, spanning both older and newer vulnerabilities. This finding exposes a critical limitation of LLMs in the context of vulnerability analysis: LLMs do not inherently distinguish individual CVEs. When queried about a specific CVE but supplied with inaccurate retrievals, the model often assumes that the provided content is correct and builds its reasoning upon it. This behavior represents a serious security risk. Even small inaccuracies in vulnerability intelligence can propagate through analysis pipelines, leading to mis-prioritization, incorrect patching decisions, or overlooked exploitation paths. Such fragility makes LLM-based systems particularly vulnerable in high-stakes cybersecurity operations, where precision and verification are non-negotiable.

This raises an important caution for the cybersecurity community: analysts should not blindly trust retrieved content when integrating LLMs into enterprise systems. Instead, agentic AI systems should adopt conservative workflows that prioritize verification of retrieved information before taking actions. Establishing a robust, verification-driven retrieval pipeline can help mitigate risk, ensuring that AI-assisted vulnerability analysis remains both accurate and trustworthy.

\subsection{LEA: verifiable LLM-assisted CVE analysis w/ valid vs. generic, vs. non-retrieval} 
\label{sec: discussion}

We validated LEA's effectiveness in modeling the distribution when different retrievals are applied, and we raised concern of the incorrect retrieval case, as they mirror the distribution of the valid retrieval. However, the main utility of LEA is in the distinction of a non-retrieval vs. generic retrieval vs. informative retrieval for vulnerabilities. In Figure \ref{fig:lea_entire} we demonstrated that for these cases, the distributions are different. 

Thus, by applying the filtering of stop-words and thresholding, we demonstrate how LEA is actually differentiating between these cases. Figure \ref{fig:results_scatter_lea} shows this distinction across the 500 CVEs over different years. As can be seen, when the retrieved context contains informative content, LEA's $A^{\rag}$ can model this dependence pretty clearly. Interestingly, This distinction is 100\% for the CVEs in 2025, and they all came after the knowledge cutoff of the tested LLMs. Out of the models, the larger ones show tendency to differentiate better between the generic vs. valid cases, while the small model (\textit{Llama3.2-3b}) tends to not over rely on some valid cases. 

To quantitatively assess the detection accuracy between these cases, we partition the dataset into a 80\% / 20\% train-test split and evaluated the $A^{\rag}$ scores against an optimal threshold determined as the point on the ROC curve closest to the perfect classifier (0,1). This point balances the trade-off between maximizing true positive detection while minimizing false positive rates. Applying this threshold ensures a clearer separation between valid, informative retrievals vs. generic or non-retrieval cases. The results, summarized in Table \ref{tab:lea_model_comparison}, show consistently strong discrimination across models: \textit{Mistral-24b} achieves a test accuracy of 0.967, \textit{Gemma3-27b} reaches 0.957, and \textit{Llama3.2-3b} achieves 0.927. These findings suggest that even comparatively smaller models can leverage LEA to reliably filter retrieval quality, while larger models demonstrate a sharper boundary between signal and noise. Operationally, integrating these thresholds into a triage pipeline will let analysts automatically prioritize CVEs that carry genuinely informative context (including post-cutoff/zero-day cases), shorten time-to-assess, and lower cognitive load by filtering out generic/non-retrieval noise.

Overall, these findings suggest that LEA not only provides a quantitative measure for assessing retrieval usefulness but also surfaces important behavioral differences across models and datasets. Its ability to highlight context-dependent vulnerability reasoning makes it a promising tool for analyzing retrieval-augmented vulnerability analysis pipelines, particularly in real-world settings where retrieval quality varies and unseen CVEs continually emerge.

\section{Conclusion}
\label{sec/conclusion}
We developed a novel, explainable metric called \textbf{LEA} to reveal the distribution of the extent LLM generated responses depend on the foundational knowledge versus retrieved context. Our analysis showed that while later layers treat tokens more independently, layer-0 exhibits dependence on retrieved inputs. Experiments on 500 CVEs over a 10-year span across multiple models revealed that LLM responses, especially for structured entities like CVE-IDs, stem from learned generalization rather than memorization. We further demonstrated its verifiable use for vulnerability analysis to show LLM generated responses are insightful by quantifying the expected distribution when properly retrieved from verifiable sources. 

Moreover, we validated LEA’s practical utility in vulnerability assessment workflows by showing its ability to reliably distinguish between informative and non-informative responses across non-retrieval, generic-retrieval, and valid-retrieval settings with more than 95\% accuracy. This capability enhances analyst confidence by quantifying the contribution of verifiable evidence in model outputs, therefore, mitigating risks of overreliance on ungrounded responses. 
As LLMs become increasingly embedded in cybersecurity operations, LEA serves as a transparent, trust-building mechanism that bridges the gap between model interpretability and operational usability. LEA not only strengthens confidence in AI-assisted vulnerability analysis but also lays the groundwork for developing more accountable and robust retrieval-augmented systems in security-critical domains.

\section{Acknowledgments}
This material is based upon work supported by the National Science Foundation under Grant No. 2344237 and No. 2228001.

\appendix

\bibliographystyle{plain}
\bibliography{REFERENCES}

\end{document}